\documentclass[journal]{IEEEtran}
\def\BibTeX{{\rm B\kern-.05em{\sc i\kern-.025em b}\kern-.08em
    T\kern-.1667em\lower.7ex\hbox{E}\kern-.125emX}}
\usepackage{cite}
\usepackage{caption}
\usepackage{subcaption}
\usepackage{graphicx}
\usepackage{amsmath}
\usepackage{amssymb}
\usepackage{amsthm}

\usepackage{stfloats}
\usepackage{bm}
\usepackage{algorithmic}
\usepackage{xcolor}
\usepackage{tikz}
\usetikzlibrary{shapes,arrows,positioning,plotmarks,shadows,calc,matrix,fit,patterns}
\usepackage{pgfplots}
\usepackage{todonotes}
\usepackage{datetime}
\usepackage{cancel}
\usepackage{hhline}
\usepackage{cases}
\usepackage{nicefrac}
\usepackage{etoolbox}
\usepackage{algorithm}
\usepackage{algorithmic}
\usetikzlibrary{plotmarks}
\usetikzlibrary{arrows.meta}
\usepackage{grffile}
\usepackage{dsfont}

\newcommand{\figref}[1]{Fig.~\ref{#1}}

\newcommand{\lemmaref}[1]{Lemma~\ref{#1}}
\newcommand{\defref}[1]{Definition~\ref{#1}}

\newcommand{\tr}{\text{tr}}
\newtheorem{proposition}{Proposition}
\newtheorem{definition}{Definition}

\newtheorem{theorem}{Theorem}
\newtheorem{lemma}{Lemma}
\newtheorem{remark}{\textit{Remark}}

\usepackage{graphicx}
\usepackage{soul}
\usepackage{amsthm}
\usepackage{siunitx}
\usepackage{bm}
\IEEEoverridecommandlockouts
\begin{document}
%Secure transmission in IRS-assisted systems: Reflecting elements selection with outdated CSI\
% Secure mm-Wave Transmission with Randomly Located Eavesdroppers: IRS and Frequency Diverse Array Assistance
%\title{{Using IRS against Nearby Eavesdropping in mmWave/THz Communication Systems}\
	\title{{Enhancing the Secrecy Rate with Direction-range Focusing with FDA and RIS}\
%\thanks{This work was funded by the Federal Ministry of Education and Research
%	(BMBF) of the Federal Republic of Germany (Förderkennzeichen 16KISK095, 6G-ANNA).}
\thanks{This work was supported by the German Federal Ministry of Education and Research (BMBF) project 6G-ANNA [grant agreement number 16KISK095].
}
}

\author{\IEEEauthorblockN{Chu Li, Stefan Roth and Aydin Sezgin}\\
	\IEEEauthorblockA{Ruhr-Universit\"at Bochum, Germany \\
		Email:  \{chu.li, stefan.roth-k21, aydin.sezgin\}@rub.de}
%	\and
%	\IEEEauthorblockN{Zhu Han }
%	\IEEEauthorblockA{University of Houston, USA\\
%		zhan2@uh.edu}
%	\thanks{C. Li and A. Sezgin are with the
%		Institute of Digital Communication Systems, Ruhr University Bochum,
%		44801 Bochum, Germany (e-mail: chu.li@rub.de;aydin.sezgin@rub.de).}
%	\thanks{M. van Delden is with the Institute of Electronic Circuits, Ruhr University Bochum,
%		44801 Bochum, Germany (e-mail:  marcel.vandelden@est.ruhr-uni-bochum.de).}
%	\thanks{Z. Han is with the Department of Electrical and Computer Engineering in the University of Houston, Houston, TX 77004 USA, and also with the Department of Computer Science and Engineering, Kyung Hee University, Seoul, South Korea, 446-701 (Email: zhan2@uh.edu).
}
%\and
%\and
%\and

\maketitle

	\begin{abstract}
%		...
One of the great potentials to improve the confidentiality in mmWave/THz at the physical layer of technical communication, measured by the secrecy rate, lies in the use of reconfigurable intelligent surfaces (RISs). However, an important open problem arises when the eavesdropper is aligned with the legitimate user or in proximity to the RIS or legitimate user. The limitation comes, on one hand, from the high directional gain caused by the dominant line-of-sight (LOS) path in high-frequency transmission, and, on the other hand, from the high energy leakage in the proximity of the RIS and the legitimate user. 
To address these issues, we employ {the concept of} frequency diverse arrays (FDA) at the base station (BS) associated with random {inverted} transmit beamforming and reflective element subset selection (RIBES). 
%unlike in most current studies where prior information of the eavesdropper is assumed to be given,
{More specifically,} we consider a passive eavesdropper {with unknown location} and design the transmit beamforming and RIS configuration based on the channel information of the legitimate user only. In this context, the secrecy rate with the proposed transmission technique is evaluated in the case of deterministic eavesdropper channel, demonstrating that we can ensure a secure transmission regarding both direction and range.  Furthermore, assuming no prior information about the eavesdropper, we describe the wiretap region and derive the \emph{worst-case secrecy rate} in closed form. The latter is further optimized by determining the optimal subset sizes of the transmit antennas and reflective elements. 
Simulations verify {the correctness of the closed-form expressions} and demonstrate that we can effectively improve the secrecy rate, especially when the eavesdropper is close to the RIS or the legitimate user.   
\end{abstract}


\section{Introduction}
{Security has always been a major concern in wireless communication systems, and it becomes even more crucial in 6G networks. In most current communication standards, security aspects are implemented on higher layers, where cryptography-based methods can be applied. 
However, the implementation of these technologies in 6G networks faces significant challenges due to the proliferation of low-resource devices and the existence of heterogeneous networks \cite{osorio2019physical}. {To complement cryptography-based methods and effectively meet the security requirements in future networks, physical layer security is envisioned as a promising solution \cite{poor2017wireless, wu2018survey, 7270404}.}
Furthermore, to enhance the performance of the physical layer security technique, reconfigurable intelligent surfaces (RISs) have emerged as a promising solution. RIS is a plain surface consisting of massive, low-cost reflective elements with configurable amplitude and phase. 
%Due to the intelligent configurable properties of RIS, it exhibits significant potentials in improving communication quality \cite{10200263}, enhancing network resilience \cite{10104574}, improving localization and sensing accuracy \cite{vcivsija2021ris,zhang2022toward, tewes2022irs}, as well as enhancing security in wireless communication and sensing \cite{lotfi2023under, 10279440, staat2022irshield}. Additionally, RIS-assisted systems can be designed to be robust against hardware impairments \cite{ 9961236, 9295369}.
Due to its configurable properties, RISs exhibit significant potentials in improving communication quality\cite{10200263,9961236, 9295369}, enhancing network resilience\cite{10104574}, advancing localization and sensing capabilities\cite{vcivsija2021ris,zhang2022toward, tewes2022irs}, and contributing to various aspects of covert communication \cite{lotfi2023under} and physical layer security\cite{10279440, staat2022irshield, 9279316, qiao2020secure, 9024636,hong2020artificial, guan2020intelligent, alexandropoulos2021safeguarding}.

The secrecy rate performance in RIS-assisted systems has been investigated in numerous recent works.
%	
%	To enhance the secrecy rate performance, 
%% of physical layer security techniques, particularly in bolstering the system's ability to resist eavesdropping attacks, 
%RIS are employed in numerous recent works.
%\cite{9279316, qiao2020secure, 9024636, hong2020artificial, guan2020intelligent, alexandropoulos2021safeguarding}. 
In \cite{9279316, qiao2020secure, 9024636}, the secrecy rate is optimized by a joint design of the transmit-beamforming and RIS phase shifts. The works \cite{hong2020artificial, guan2020intelligent, alexandropoulos2021safeguarding} additionally utilize artificial noise to improve the secrecy rate in an RIS-assisted system. 
However, the approaches proposed in the aforementioned works to optimize the secrecy rate assume the channel state information of the legitimate user and the eavesdropper are uncorrelated and independent.
%an {independent channel} between the base station (BS) and the legitimate user regarding eavesdropper.
 Under such assumption, the transmit beamforming and RIS are jointly designed to enhance the received signal at the legitimate receiver while simultaneously weaken the received signal at the eavesdropper, thus improving secrecy. However, these approaches become ineffective in mmWave/THz communication systems, if the eavesdropper is spatially aligned with the legitimate user or in proximity to the legitimate user or RIS. In this case, the channels of the legitimate user and the eavesdropper are highly correlated, primarily due to the dominant line-of-sight (LOS) path. Moreover, compared to transmission in low and moderate frequencies, the channel gain in mmWave/THz attenuates more rapidly with an increasing propagation path. {Note that an eavesdropper} located close to the RIS may experience a {significantly} larger channel gain compared to that of the legitimate user. {Hence, an eavesdropper at this location} may have the ability to decode the transmitted signal correctly. 

{To decorrelate the channels of receivers in close proximity to each other, frequency diverse array (FDA) has been proven  as a promising technique in many recent studies \cite{8081593,8078202,cheng2021physical,10097703}.
FDA was initially proposed in \cite{1631800}, subsequently gaining significant attention in radar applications \cite{farooq2007application, xu2015joint, wang2016overview, 6732922}.  
By exploiting a minor frequency shift across the antennas, FDA can generate a direction-range-dependent beam pattern. Thus, it can be also used for improving the physical layer security performance in terms of secrecy rate and covert rate in mmWave/THz transmission
\cite{8078202, 10097703, cheng2021physical,  7817778,9133276, li2023covert}.}
%In particular, the authors in \cite{8078202,10097703} proposed joint optimization of frequency shifts and transmit-beamforming to maximize the secrecy rate, given the channel information of the legitimate user and eavesdropper. They demonstrated the advantages of FDA beamforming, particularly in cases where the eavesdropper is in proximity to the legitimate user.
%A random FDA-based directional modulation with artificial noise is designed to ensure secure transmission in \cite{7817778,9133276,li2023covert}, where the ergodic secrecy capacity and a lower bound for the secrecy capacity are derived.
%To achieve comprehensive security, the frequency shift at each antenna needs to be carefully designed \cite{8078202,10097703, cheng2021physical} or utilize a randomly distributed frequency shift across the antennas \cite{7817778,9133276,li2023covert}, which may increase the circuit cost at the transmitters. 
%However, all the approaches proposed in the aforementioned works require accurate channel information of the eavesdropper, which is challenging to obtain in practice. 
More specifically, by solving {different} non-convex optimization problems, the frequency shift at each antenna is carefully designed to optimize the secrecy rate in \cite{8078202, cheng2021physical} and the covert rate in \cite{10097703}. In these works, the optimization problem is solved by assuming either full or partial information about the eavesdropper, which is challenging to obtain in practice. Alternatively, random FDA is utilized to enhance the secrecy rate in \cite{7817778, 9133276} and covert rate in \cite{li2023covert}. However, such approach may lead to an increase in the circuit cost at the transmitters. Moreover, direct transmission is prone to blockages in mmWave/THz. To address this issue, an RIS can be employed to {establish an additional strong communication path} between transceivers \cite{9759225}. {Different to the above works, we investigate the secrecy performance in an RIS-assisted system with FDA in the presence of a passive eavesdropper.}

%\textcolor{orange}{Thus, it is worthwhile to investigate the secrecy performance in an IRS-assisted system with FDA, which has not been studied in state-of-the-art works.}
%
%
%In this study, we consider an IRS-assisted communication system in the presence of a passive eavesdropper.

{Furthermore, the eavesdropper can position itself in any of various vantage locations to intercept confidential messages.} Therefore, it is crucial to develop an approach that safeguards the transmission in all circumstances. To resist {against an eavesdropper in} any undesired directions, the works \cite{6544472, 7386629} utilize {a random transmit antenna switching technique}. In this approach, the transmitted signal is intentionally degraded in any undesired direction, such that the overall secrecy performance is enhanced. More specifically, a random antenna subset is selected for each transmit symbol to perform beamforming, while the remaining antennas are turned off. Additionally, a new random selection is generated for each following transmit symbol. {However, such kind of sparse antenna array configuration provides an opportunity for a smart eavesdropper capable of conducting measurements at different locations to intercept and recover the transmitted signal \cite{7178504}. In addition, the use of switched antenna technique requires a switch at each antenna that controls the ON-OFF status.}
Subsequently, random inverted transmit antennas are utilized in \cite{8329405, 7876781, 9387394, hong2020fixed}, also referred to as randomized radiation or
randomized beamforming. {In this technique, a random subset of the transmit antennas is selected for conventional beamforming, while the inverted beamforming, i.e., conventional beamforming with an additional weight of $e^{j \pi} = -1$, is performed at the remaining antennas.}
% where beamforming is also performed at the remaining antennas with \textcolor{blue}{an additional weight of $e^{j \pi} = -1$}. 
%Subsequently, random inverted transmit antennas are utilized in \cite{8329405, 7876781, 9387394, hong2020fixed}, also referred to as randomized radiation or randomized beamforming. In this technique, a random subset of the transmit antennas is selected to perform conventional beamforming, while inverted beamforming, which involves an additional weight of $e^{j \pi} = -1$ to the conventional beamforming, is applied to the remaining antennas.
% In other words, conventional beamforming is applied with a random subset of antennas, and inverted beamforming, which involves an additional weight of $e^{j \pi} = -1$, is applied to the rest
 As all antennas are active during the transmission, switching between states is not needed. {Furthermore, the work \cite{10279440} proposed reflective element subset selection (ESS) in RIS-assisted secure systems, where a subset of the reflective elements is selected to enhance the secrecy rate.}
%
%
%	\textcolor{pink}{In general, two selection schemes can be applied: random switched antenna selection and random inverted antenna selection \cite{8329405}. With the random switched antenna selection, only the selected antennas are used to transmit the data, while the remaining antennas are inactive. This sparse antenna array configuration provides an opportunity for a smart eavesdropper capable of conducting measurements at different locations to intercept and recover the transmitted signal \cite{7178504}. In addition, the use of switched antenna technique requires a switch at each antenna and at each reflective element that controls the ON-OFF status. The latter technique does not require a switch, and all antennas and reflective elements are utilized at each symbol interval. More specifically, this technique can be easily applied by introducing a weight vector containing only 1 and -1 to the conventional transmit-beamforming vector. Furthermore, the weight of -1 can also be easily implemented at each IRS element by introducing an additional phase shift of $\pi$ to the optimal phase shift, as $e^{j \pi} = -1$. Therefore, we apply random inverted antenna selection, referred to as RIAS, in this work.} 

$\mathit{Contribution}$:
{In this work, we develop a transmission technique for RIS-assisted systems that ensures secure communication regarding direction and range, without requiring prior information about the eavesdropper. To this end,} we propose random inverted transmit beamforming and reflective element subset selection (RIBES), to be utilized jointly with a FDA at the BS. More specifically, we investigate the secrecy rate performance in mmWave/THz systems, where a BS communicates with a legitimate receiver Bob, in the presence of a passive eavesdropper Eve.  
%	 By the proposed algorithm, the signal received in any undesired location in randomized.}
  We demonstrate that security in terms of confidentiality can be achieved through the proposed transmission technique.} This holds even with a simple linear FDA, where the frequency shift between neighboring antennas remains the same. 
%{To the best of our knowledge, this is the first joint investigation of FDA and \textcolor{blue}{RIBES} in an RIS-assisted system.} 
%\textcolor{orange}{To the best of our knowledge, this is the first time that investigates FDA and RIAS jointly in an IRS-assisted system.} 
%\textcolor{green}{To the best of our knowledge, this is the first time that the benefits of FDA and RIAS are jointly utilized in an IRS-assisted system.} 
 The main contributions are as follows: 
\begin{itemize}
	\item We employ FDA at the BS with {RIBES}, considering the presence of a passive eavesdropper.  In this context, we design the transmit beamforming and RIS configuration to maximize the received SNR at {Bob}.   
	\item Applying the proposed transmit beamforming and RIS configuration, {we define the scaling factor of the received signal and derive its statistical properties,} laying the foundation for analyzing secrecy performance. Afterwards, we analyze and derive
	closed-form expressions for the secrecy rate, both with and without employing FDA and {RIBES}. 
	\item We formulate a definition of the wiretap area for the proposed transmission technique. Furthermore, assuming Eve is located within the wiretap area, we derive the upper bound of the received SNR at {Eve}.
	\item Based on the above derivations, {we find a closed-form expression for the optimal subset sizes of the transmit antennas and reflective elements that maximize the \emph{worst-case secrecy rate}.}
%	Based on the above derivations, we find a closed-form expression for the number of selected transmit antennas and reflective elements that optimize the worst-case secrecy rate. 
	In addition, we simplify the expressions for the case where the received SNR at {Bob} is significantly larger than \SI{0}{dB}. 
	\item We determine the optimal frequency increment that maximizes the secrecy performance given the channel information of Eve. The resulting secrecy rate is considered as an upper bound in this work.
\end{itemize}

$\mathit{Organization}$:
The rest of the paper is organized as follows. We introduce the system model in {Sec.}~\ref{sec:chmod}. Afterwards, the proposed transmit beamforming and RIS configuration with {RIBES} are provided in Sec.~\ref{sec:beamforming_design}. Subsequently, we analyze the secrecy rate with and without the use of FDA and {RIBES} in Sec.~\ref{sec:secrecy}. Then, the proposed {RIBES} algorithm is optimized in Sec.~\ref{sec:optimization}. We present the theoretical results corroborated with the simulation results in Sec.~\ref{sec:numerical_results}. Finally, Sec.~\ref{sec:conclusion} concludes the paper. 

%From the eavesdropper's perspective, to successfully intercept confidential messages, they may position themselves close to the IRS or the legitimate user. In such cases, even with prior information about the eavesdropper, the proposed algorithms in the aforementioned works may not be effective, especially in mmWave/THz communication.
%The system model is given in Sec.~\ref{sec:sysmod}. We introduce the LMMSE channel estimation algorithm in Sec.~\ref{sec:CE}. We analyze the downlink performance of an IRS-assisted system in Sec.~\ref{sec:dl}. Simulation results are corroborated with the analytical results in Sec.~\ref{sec:results}. Finally, Sec.~\ref{sec:conclusion} concludes the paper. 

$\mathit{Notation}$: Throughout this paper, boldface lower and upper case symbols are used to denote the vectors and matrices, respectively.  $(\cdot)^H$ and  $\circ$   represent the  Hermitian transpose operator and Hadamard product. The notation ${ \mathbf{X}}_{i,j}$ indicates the element at position $(i, j)$ of the $\mathbf{X}$.  $[x]^+$ is a short form for $\max(0,x)$. {The integer closest to $x$ is denoted as $ \left\lfloor x  \right\rceil$.}
The expectation and variance operator are denoted as $\mathbb{E}\left[ \cdot\right]  $  and $\mathbb{V}\left[ \cdot\right] $, respectively.

\section{System Model}
\label{sec:chmod}
\begin{figure}
	\centering
	\includegraphics[width=1\linewidth]{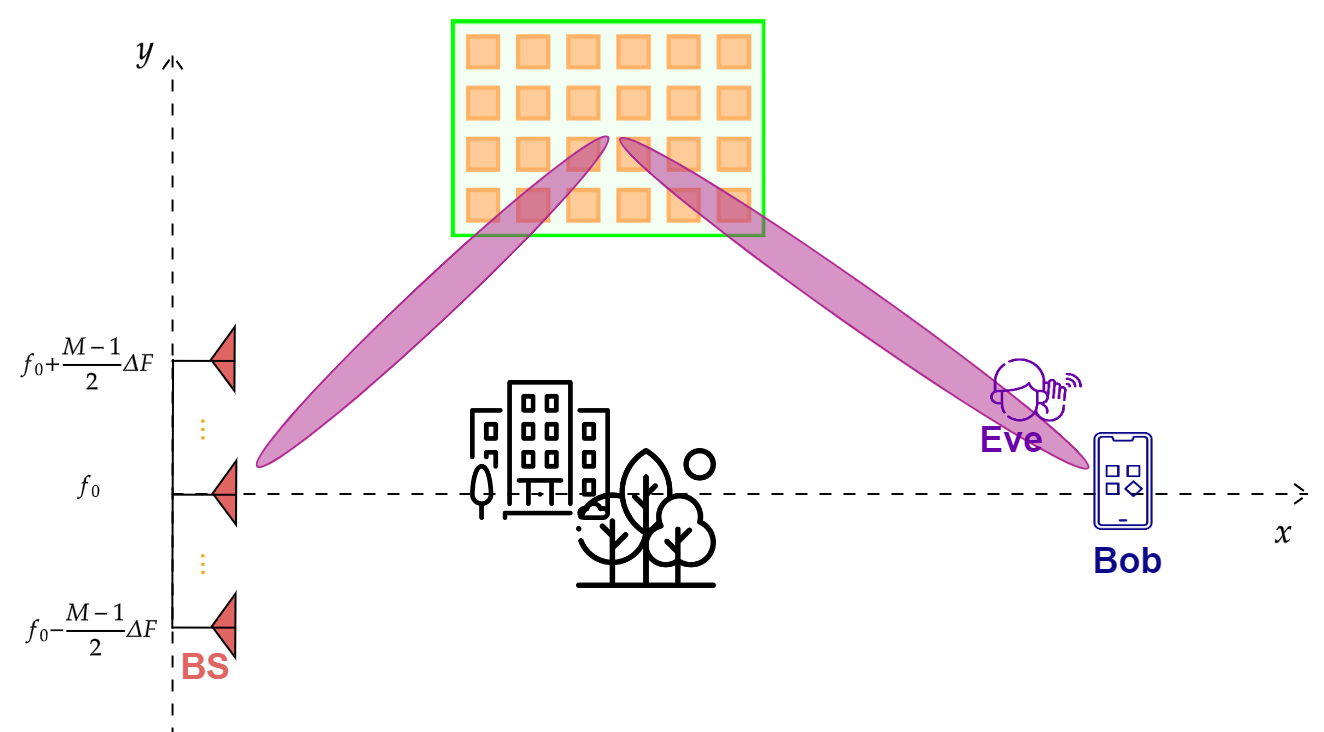}
	\caption{{RIS-assisted secure transmission with FDA at the BS in the presence of an eavesdropper}} \vspace{-1em}
	\label{fig:systemmodel}
\end{figure}
%\begin{figure}
%	\centering
%	\includegraphics[width=0.75\linewidth]{senario2}
%	\caption{IRS-assisted system with a legitimate user (Bob) and an eavesdropper (Eve)}
%	\label{fig:senario}
%\end{figure}
%\subsection{FDA at the BS}
We consider RIS-assisted {secure (confidential)} transmission in a mmWave/THz system as shown in \figref{fig:systemmodel}, where a $M$-antenna BS is communicating with a single-antenna legitimate receiver, Bob, in the presence of a passive eavesdropper equipped with a single antenna, Eve. Furthermore, we consider a two-dimensional area, {where the BS has equally spaced antennas and is positioned at the origin.} 
%The BS antennas are uniformly distributed along the y-axis, with a separation of $d_{\mathrm{BS}}$. 
We assume that direct transmission from the BS to all users is obstructed by blockage. To overcome this blockage, {we employ an RIS positioned at $[x_{\text{RIS}}, y_{\text{RIS}}]$ to establish LOS links between the BS and the RIS, as well as between the RIS and the receivers.} In addition, the RIS consists of a total of $N = N_\mathrm{V} \times N_\mathrm{H}$ reflecting elements, where $N_\mathrm{V}$ and $N_\mathrm{H}$ represent the number of elements per column and row, respectively. In addition, we assume all components of the RIS-assisted system are situated in the far field.

In this work, we utilize FDA at the BS to establish a {secure} transmission with both direction and range protection. In this particular configuration, the radiation frequency at the $m$-th antenna is written as 
\begin{align}
	f_m &= f_0 + \Delta f_m \nonumber \\
	&=f_0 + \left( m-\frac{M+1}{2}\right) \Delta F,	\quad m \in \{1,\dots, M\},
\end{align}	
where $f_0$ is the center frequency, $\Delta f_m$ represents the frequency shift at $m$-the transmit antenna, and $\Delta F$ denotes the frequency increment between adjacent antennas. Furthermore, in this study, we assume that the frequency shift  $\Delta f_m $ is significantly smaller than the center frequency.  More specifically, we impose a constraint on the maximal frequency shift, i.e.,  $\max_m(\Delta f_m) < 10^{-3} f_0$. 

{It is noted that we are considering mmWave/THz transmission, where the LOS component dominates over the NLOS components due to high spreading, reflection, and scattering losses\cite{8078202 }. Therefore, in this work, we exclusively focus on the dominant LOS component and model the mmWave/THz channel according to the geometry channel model with one dominant component \cite{ 9295369, 6732923, 7876781}. }
%	investigate the benefits of the proposed technique. Furthermore, the path loss of the mmWave/THz channel is modeled as
%   \begin{align} 
%   %	L (R) = 60 + 20 \log_{10}(R) [\mathrm{dB}],
%   	L(R) = L_0 + 10 \alpha \log_{10}(R) \ [\mathrm{dB}],
%   \end{align}	
%where $R$, $L_0$ and $\alpha$ are the propagation distance, path loss value at a reference distance of \SI{1}{\meter}, and path loss exponent. Here, we employ a large reference path loss $L_0 = \SI{60}{dB}$ to align with the propagation conditions of mmWave/THz. Additionally, we set $\alpha$ to 2. } 
More specifically, we use $\mathbf{G} \in \mathbb{C} ^{M \times N}$ to represent the BS-RIS channel, where the element of $\mathbf{G}$ at position $(m,n)$ is given by
\begin{align}
	\label{eq:BS_IRS_ch}
	\mathbf{G}_{m,n} = \sqrt{L_\mathrm{G}(R_1)}e^{-j {2\pi f_m}\tau_{m,n}^{\text{BS-RIS}}},
\end{align} 
in which $L_\mathrm{G}(R_1)$ is the path loss with $R_1$ being the distance between the BS and RIS.
Further, $\tau_{m,n}^{\text{BS-RIS}}$ in \eqref{eq:BS_IRS_ch} is the propagation delay from the $m$-th transmit antenna to the $n$-th RIS element, and it is computed as	 
\begin{align}
	\label{eq:delay_BS-IRS}
	\tau_{m,n}^{\text{BS-RIS}} = \frac{1}{c}\bigg( R_{1}&-\left( m-\frac{M+1}{2}\right) d_{\mathrm{BS}} \sin(\theta_{\mathrm{tx}}) \nonumber \\ &+\left( n_\mathrm{V} - \frac{N_\mathrm{V}+1}{2}\right) d_\mathrm{V} \sin(\theta_{\mathrm{tx}})\nonumber \\ &+\left( n_\mathrm{H} - \frac{N_\mathrm{H}+1}{2}\right)d_\mathrm{H} \cos(\theta_{\mathrm{tx}}) \bigg).
\end{align}
Here, $n_\mathrm{H}= \mod \left(n, N_{\mathrm{H}}\right)$ and  $n_\mathrm{V}=\left\lfloor n / N_{\mathrm{H}}\right\rfloor$ are the horizontal and vertical indices of the $n$-th element, where $n \in \{ 1, \cdots, N\}$, respectively. {In addition, $d_{\mathrm{BS}}$ represents the antenna separation at the BS, while $d_\mathrm{H}$ and $d_\mathrm{V}$ denote the horizontal width and vertical height of each reflective element, respectively.} Furthermore, we consider the setup $d_{\mathrm{BS}} = d_\mathrm{H}= d_\mathrm{V} = \frac{c}{2 f_0}$, {where $c$ denotes the speed of light.}  Moreover, we have $\theta_{\mathrm{tx}} = \arcsin\left(\frac{y_{\text{RIS}}}{R_1}\right)$ representing the angle of departure (AOD) at the BS. 
%Note that, as we consider a 2-dimensional geometric scenario with the BS located on the y-axis, the AOD at the BS is equal to the angle of arrival (AOA) at the IRS.
Note that, within the considered two-dimensional geometric scenario, the AOD at the BS is
equal to the angle of arrival (AOA) at the RIS.
Furthermore, since multiple frequencies are utilized at the BS, we represent the channel between RIS and the receiver using an $M \times N$ matrix, i.e., $\mathbf{H}_{\mathrm{UE}}$, {where the $m$-th column corresponds to the channel at frequency $f_m$.} It is noted that $ \mathrm{UE} \in \{\mathrm{B}, \mathrm{E}\}$ can represent either Bob (the legitimate receiver) or Eve (the eavesdropper).  Similar to \eqref{eq:BS_IRS_ch}, we have
%the $(m,n)$-th element of IRS-UE channel is given by
%\begin{align}
%	\label{eq:BS_UE_ch}
%	\mathbf{H}_{m,n} = \sqrt{\frac{C_0}{R_{\text{UE}}^{\alpha_2}}}e^{-j {2\pi f_m}\tau_{n}^{\text{IRS-UE}}},
%\end{align}
\begin{align}
	\label{eq:BS_UE_ch}
	\mathbf{H}_{\text{UE}, {(m,n)}} = \sqrt{L_\mathrm{H}(R_{\text{UE}})}e^{-j {2\pi f_m}\tau_{n}^{\text{RIS-UE}}},
\end{align}
%where $R_{\text{UE}}$ represents the distance between the IRS and the respective user, and $\alpha_2$ is the path loss exponent of the IRS-UE channel. 
where $L_\mathrm{H}(R_{\text{UE}})$ is the path loss of RIS-UE channel, and  $R_{\text{UE}}$ represents the distance between the RIS and the respective receiver.
The propagation delay from the $n$-th reflective element to the receiver $\tau_{n}^{\text{RIS-UE}}$  can be computed as
\begin{align}
	\label{eq:delay_IRS-UE}
	\tau_{n}^{\text{RIS-UE}} = \frac{1}{c} \bigg( R_{\text{UE}} &+\left( n_\mathrm{V}- \frac{N_\mathrm{V}+1}{2}\right) d_\mathrm{V} \sin(\theta_{\text{UE}})\nonumber \\ &-\left( n_\mathrm{H}- \frac{N_\mathrm{H}+1}{2}\right)d_\mathrm{H} \cos(\theta_{\text{UE}}) \bigg),
\end{align}	
in which $\theta_{\text{UE}}$ is the AOA at the receiver. Given that the receiver's location is specified as $[x_{\text{UE}},y_{\text{UE}}]$, $\theta_{\text{UE}}$ can be computed by $\arccos\left( \frac{x_{\text{UE}}-x_{\text{RIS}}}{R_{\text{UE}}}\right) $.  
%\subsection{Received Signal}

In addition, we use $x(k) \in \mathbb{C}$ to represent the transmitted data symbol with unit variance, where $k$ denotes the symbol index. Also, the transmit beamforming vector at the BS is represented by $\boldsymbol{w}(k) \in \mathbb{C}^{M\times 1}$. Thereby, the received signal can be expressed as 
\begin{align}
	\label{eq:received_signal_Bob}
	y_{\text{UE}}(k) =\sqrt{P}\mathbf{h}_{\text{UE}}^H \boldsymbol{w}(k) x(k) + n_{\text{UE}}(k), 
\end{align} 
%  \quad \text{UE} \in [\text{Bob}, \text{Eve}],
%Let $\boldsymbol{w} \in \mathbb{C}^{M\times 1}$ and $x$ represent the transmit beamforming vector at the BS and the transmit symbol, respectively. The received signal observed at Bob can be expressed as
%\begin{align}
%	\label{eq:received_signal_Bob}
%	y_{\text{Bob}} =\mathbf{h}_{\text{Bob}}^H \boldsymbol{w} x + n_{\text{Bob}},
%\end{align}
where $P$ is the transmit power, { $n_{\text{UE}}(k)$ is the thermal noise at the receiver, distributed as $n_{\text{UE}}\sim \mathcal{CN}(0, \sigma^2_{\text{UE}})$,} and $\mathbf{h}_{\text{UE}} \in \mathbb{C}^{M \times 1}$ is the cascaded channel given by
\begin{align}
	\label{eq:cascaded_ch_Bob}
	\mathbf{h}_{\text{UE}} =   ( \mathbf{G} \circ \mathbf{H}_{\text{UE}}) \mathbf{v}.
\end{align}
Here, $\mathbf{v}$ is the RIS vector, which can be expressed as
\begin{align}\mathbf{v} = [e^{j \phi_1},  \ldots, e^{j \phi_n}, \ldots, e^{j \phi_N}]^T, 
\end{align}
where $\phi_n$ represents the phase shift of the $n$-th reflective element. The phase shifts can be dynamically adjusted by the BS. 
{Thereby, we focus on a frequency-flat RIS configuration, i.e., the applied phase shifts are the same across all frequencies. This kind of configuration is utilized in} the majority of studies that have explored multiple frequencies in RIS-assisted transmission \cite{8937491, 9740451, 9133184}. 
%Within frequency-flat IRS, the applied phase shifts are the same across all frequencies.
 Note that, frequency-selective phase shifts within an RIS can be realized by carefully designing the circuit topology of each reflective element \cite{9514409}. However, such a design increases the hardware complexity.  

	\section{Transmit beamforming and RIS design}
	\label{sec:beamforming_design}
We utilize the secrecy rate to evaluate the performance of the proposed method. Let $\gamma_\mathrm{B}$ and $\gamma_\mathrm{E}$ represent the received SNR at Bob and Eve, respectively. The secrecy rate, as defined in \cite{wyner1975wire}, can be expressed as 
	\begin{align}
		\label{eq:sec_cap}	
		C_s(\gamma_{\mathrm{B}},\gamma_{\mathrm{E}}) = \left[ \log_2 (1+\gamma_{\mathrm{B}})-\log_2 (1+\gamma_{\mathrm{E}})\right] ^+.
	\end{align}
	Under the assumption of a passive eavesdropper, we design the transmit beamforming vector and RIS solely utilizing {the value of the BS-to-Bob channel}. In this work, we assume that perfect information of {the BS-to-Bob channel} is {given at} the BS. {As a result, the optimal beamformer and RIS configuration that maximize the secrecy rate are equivalent to those that maximize the received SNR at Bob. } In the following, we first introduce the conventional technique for optimizing transmit beamforming and RIS, and then we introduce the transmit beamforming and RIS design with RIBES. 

	\subsection{Conventional transmit beamforming and RIS design }
	\label{sec:3_A}
	 %$\bm{h}$
	%
	%%P1 is equivalent to the objective of maximizing the received SNR at Bob, i.e., $\gamma_{B}$.
	%In this section, we begin by introducing the conventional technique for optimizing transmit beamforming and IRS. Within this framework, it
	Maximizing the received SNR in a multiple-input single-output (MISO) system is a well-established problem \cite{tse2005fundamentals}. In this context, the optimal strategy for transmit beamforming is maximum ratio transmission (MRT), which is
	\begin{align}
		\label{eq:precoding}
		\boldsymbol{w} = \frac{\mathbf{h}_{{\mathrm{B}} }}{\left\| \mathbf{h}_{{\mathrm{B}} }\right\| },
	\end{align}
	where, according to \eqref{eq:cascaded_ch_Bob}, $\mathbf{h}_{{\mathrm{B}}} = ( \mathbf{G} \circ \mathbf{H}_{{\mathrm{B}}}) \mathbf{v} $. From this, the received SNR at Bob can be expressed as 	 \begin{align}
		 	\label{eq:gamma_B_1}		
		 	\gamma_\mathrm{B }
		 	= \frac{P}{\sigma_{\mathrm{B}}^2}{\left| \mathbf{h}_{\mathrm{B}}^H  \boldsymbol{w}  \right| ^2} = \frac{P}{\sigma_{\mathrm{B}}^2}\tr (\mathbf{\Psi } \mathbf{V}), 
\end{align}
where $\mathbf{V} = \mathbf{v}  \mathbf{v}  ^H$ {is a rank-one matrix,}  and $\mathbf{\Psi } = ( \mathbf{G} \circ \mathbf{H}_{\mathrm{B}})^H ( \mathbf{G} \circ \mathbf{H}_{\mathrm{B}}) $. The semidefinite relaxation (SDR) technique can be applied to find the optimal RIS configuration that maximizes $\gamma_{\mathrm{B}}$ \cite{wu2019intelligent}. By employing the above, the corresponding problem is formulated as
	\begin{align}
		\label{eq:sdp}
		\max_{\mathbf{V}} \quad &\tr (\mathbf{\Psi } \mathbf{V}),  \\
		 \mathrm{s. t.} \quad &\mathbf{V}(n,n) =1, \quad n = 1, \cdots, N, \nonumber \\
		&  \mathbf{V} \succeq 0. \nonumber
	\end{align}
	It's worth noting that the rank-one constraint of $\mathbf{V}$ has been removed in order to obtain a convex problem.
	%solving \eqref{eq:sdp} cannot guarantee the rank-one constraint of $\mathbf{V}$.
	 However, solving \eqref{eq:sdp} does provide a useful upper bound for $\mathbf{\gamma}_{\mathrm {B}}$. In this work, we use this upper bound to evaluate the performance of the closed-form solution proposed in the following.

We now derive the closed-form solution for $\mathbf{v}$ that maximizes the received SNR at Bob. Therefore, we first rewrite $\mathbf{h}_{\text{UE}}$ in \eqref{eq:cascaded_ch_Bob} with two auxiliary parameters $ \tilde{\tau}_{m} $ and $\tilde{\tau}_{n}^{\text{UE}}$. 
By substituting \eqref{eq:BS_IRS_ch},  \eqref{eq:delay_BS-IRS}, \eqref{eq:BS_UE_ch} and \eqref{eq:delay_IRS-UE} into \eqref{eq:cascaded_ch_Bob}, the $m$-th element of $\mathbf{h}_{\text{UE} }$ is expressed as\vspace{-0.5em}
	\begin{align}
		\label{eq:h_UE_m}
	\mathbf{h}_{\text{UE} ,m} &=  \sqrt{L_\mathrm {G}(R_1) L_\mathrm {H}(R_{\text{UE}})} \sum_{n = 1}^{N}	e^{-j {2\pi f_m} \left( \tau_{m,n}^{\text{BS-RIS}}+ \tau_{n}^{\text{RIS-UE}}  \right) }  e^{j \phi_n} \nonumber \\
	 & =  \sqrt{L_\mathrm {G}(R_1) L_\mathrm {H}(R_{\text{UE}})} e^{-j2\pi f_m \left(  \frac{R_1+R_{\text{UE}}}{c} + \tilde{\tau}_{m} \right) }  \nonumber \\ 
	 & \quad \times 
	 \sum_{n = 1}^{N} e^{-j 2\pi f_m \tilde{\tau}_{n}^{\text{UE}}} e^{j \phi_n},
\end{align}
	where \vspace{-0.5em}
	\begin{align}
		\tilde{\tau}_{m} &= \frac{1}{c}\left( m- \frac{M+1}{2}\right) d_{\mathrm{BS}} \sin(\theta_{\mathrm {tx}}),   \\ \tilde{\tau}_{n}^{\text{UE}} &=   \frac{1}{c} \bigg( \left( n_\mathrm{V} - \frac{N_\mathrm{V}+1}{2}\right) d_\mathrm{V} \left( \sin(\theta_{\mathrm {tx}})+ \sin(\theta_{\text{UE}})\right) \nonumber \\ & +\left( n_\mathrm{H} - \frac{N_\mathrm{H}+1}{2}\right)d_\mathrm{H} \left( \cos(\theta_{\mathrm {tx}})-\cos(\theta_{\text{UE}})\right)  \bigg).
	\end{align}
	Thus, the received SNR at Bob with MRT transmit beamforming can be computed as
	\begin{align}
		\gamma_\mathrm {B} &=
		\frac{P}{\sigma_{\mathrm {B}}^2} {\left| \mathbf{h}_{\mathrm {B}}^H  \boldsymbol{w}  \right| ^2}
		=  \frac{P}{\sigma_{\mathrm {B}}^2}  \sum_{m = 1}^{M}  \left| \mathbf{h}_{\mathrm {B},m }  \right|^2   \nonumber  \\
	 =&  \frac{P}{\sigma_{\mathrm {B}}^2}L_\mathrm{G}(R_1) L_\mathrm{H}(R_{\mathrm {B}})  \sum_{m = 1}^{M} \bigg|   \sum_{n = 1}^{N}  e^{j {\phi}_n}	 e^{-j 2 \pi (f_0 + \Delta f_m){\tilde{\tau}^{\mathrm {B}}_n}} \bigg| ^2. \nonumber
	\end{align}
{Here, it is noted that the optimal RIS phase shift ${\phi}_n$ should be designed to mitigate the term $2 \pi (f_0 + \Delta f_m){\tilde{\tau}^{\mathrm {B}}_n}$. However, given our assumption of a frequency-flat configuration at the RIS, it is impossible to find a closed-form solution for the optimal phase shift to fully mitigate this term. Therefore, we design the phase shift at the RIS as }
	\begin{align}
		\label{eq:opt_IRS_closed_form}
		{\phi}_n^{\mathrm{opt}} = {2 \pi f_0 {\tilde{\tau}^{\mathrm{B}}_n}},
	\end{align}
	with which the SNR at Bob is computed as
	\begin{align}
		\label{eq:SNR_Bob_FDA}
		\gamma_{\mathrm{B}}&=\frac{P}{\sigma_{\mathrm{B}}^2} L_\mathrm{G}(R_1) L_\mathrm{H}(R_{\mathrm{B}})
		\sum_{m = 1}^{M} \bigg|  \sum_{n = 1}^{N} e^{j {\phi}_n^{\mathrm{opt}} }	e^{-j 2 \pi (f_0+\Delta f_m) {\tilde{\tau}^{\mathrm{B}}_n}} \bigg| ^2 
%		& = \frac{P}{\sigma_{\mathrm{B}}^2} L_\mathrm{G}(R_1) L_\mathrm{H}(R_{\mathrm{B}})
%		\sum_{m = 1}^{M} \bigg|  \sum_{n = 1}^{N} 	e^{-j 2 \pi \Delta f_m {\tilde{\tau}^{\mathrm{B}}_n}} \bigg| ^2
		\nonumber \\& {\leqslant}  \frac{P}{\sigma_{\mathrm{B}}^2} L_\mathrm{G}(R_1) L_\mathrm{H}(R_{\mathrm{B}}) M N^2.
	\end{align} 
{The upper bound is achieved when $\Delta f_m = 0$. In fact, as we assume a far-field scenario and $\Delta f_m \ll f_0$, the term ${2 \pi \Delta f_m \tilde{\tau}^{\mathrm{B}}_n}$ is significantly smaller compared to ${2 \pi \Delta f_0 \tilde{\tau}^{\mathrm{B}}_n}$. Consequently, the gap between the upper bound and $\gamma_\mathrm{B}$, when employing ${\phi}_n^{\mathrm{opt}}$, becomes almost imperceptible.}
%	\begin{remark}
%		Recall that, we assume a frequency-flat configuration at the IRS. }Hence, finding the closed-form solution for the optimal phase shift to completely mitigate the phase shift introduced by the FDA is challenging. Nonetheless, 
%The approxiamtion is due to our assumption of a far-field scenario and $\Delta f_m \ll f_0$, such that we can disregard the remaining term  in \eqref{eq:SNR_Bob_FDA}, as ${2 \pi \Delta f_m \tilde{\tau}^{\mathrm{B}}_n}$  is significantly smaller compared to ${ 2 \pi \Delta f_0 \tilde{\tau}^{\mathrm{B}}_n}$.
%		\textcolor{orange}{\st{The perfect alignment between the simulation results and the approximation in} \eqref{eq:SNR_Bob_FDA} validates this observation (see Sec. \ref{sec:numerical_results}).} 
%Furthermore, to evaluate the performance of the closed-form solution in \eqref{eq:opt_IRS_closed_form}, we use the upper bound obtained by solving \eqref{eq:sdp}. 
		As a closed-form solution leads to tractable secrecy performance and has significantly lower complexity compared to the SDR technique, we utilize the closed-form solution of the RIS for the remainder of the work.
\vspace{-1em}	
	\subsection{{Random inverted transmit beamforming and reflective element subset selection (RIBES)}}
	\label{sec:3B}
	By employing the conventional MRT beamforming in \eqref{eq:precoding} and the RIS configuration proposed in \eqref{eq:opt_IRS_closed_form}, we can effectively improve the received SNR at Bob, consequently leading to an enhancement in the secrecy rate. 
	However, the existence of high-energy signal powers in the proximity of Bob or the RIS may provide an opportunity for a sensitive eavesdropper to correctly decode the transmitted symbols \cite{naeem2023security}.
{To address this concern, we utilize RIBES with FDA, where subsets of the transmit antennas and reflective elements are jointly selected at each symbol interval to execute the conventional MRT beamforming and RIS configuration proposed in \eqref{eq:precoding} and \eqref{eq:opt_IRS_closed_form}, respectively.  Simultaneously, inverted transmit beamforming and inverted RIS configuration are performed on the remaining subset of the transmit antennas and reflective elements, respectively. More specifically, a subset of the antenna indices, i.e., $\{1,\dots,M\} $,  with a size of $M_s$, is randomly selected at each transmit symbol. In addition, we use $\mathcal{M}_s(k)$ to denote selected subset of the antenna indices. In this context,  
	the transmit beamforming is given by 
\begin{align}
	\label{eq:precoding_selection}
	\boldsymbol{w}^{\star}(k) =  \frac{ {\mathbf{h}}_{\mathrm{B} }^\star(k)}{\big\| {\mathbf{h}}_{\mathrm{B} }^\star(k)\big\| } \circ \mathbf{a}(k),
\end{align}
where $ \mathbf{a}(k) $ is a $M$-dimensional vector with  $a_m(k) = 1$ if $m \in \mathcal{M}_s(k)$, otherwise $a_m(k)=-1$.} 
Further, ${\mathbf{h}}_{\mathrm{B}}^\star(k)$ in \eqref{eq:precoding_selection} represents the cascaded channel of Bob with {RIBES}.
%Thereby, ${\mathbf{h}}_{\mathrm{B}}^\star(k)$ in \eqref{eq:precoding_selection} represents the cascaded channel of Bob with RIAS.  Moreover, $\mathbf{a}(k) $ is a $M$-dimensional vector with  $a_m(k) = 1$, if $m \in \mathcal{M}_s(k)$, otherwise $a_m(k)=-1$. 
%
%containing the elements $\{1,-1\}$.  At each symbol interval, a subset of the antenna indices, i.e., $\{1,\dots,M\} $,  with a size of $M_s$ is randomly selected. Here, we use $\mathcal{M}_s(k)$ to denote selected subset. 
%
% Moreover, $\mathbf{a}(k) $ is a $M$-dimensional vector containing the elements $\{1,-1\}$, where each element $a_m(k)$, $m\in\{1,\dots,M\} $,  changes randomly at each symbol interval. 
%, with individual elements denoted as $a_m (k)$, for $m\in\{1,\dots,M\} $. 
%In addition, we use $\mathcal{M}_s(k)$ to denote the set of the indices where $a_m(k)=1$, and $\left| \mathcal{M}_s(k) \right| = M_s$.
%
%
%During the transmission, $M_s$ elements of $\mathbf{a}(k)$ are set to $1$, while the remaining elements are set to $-1$. Furthermore, we use $\mathcal{M}_s(k)$ to denote the set of the indices $m$ where $a_m(k)=1$.
Meanwhile, the phase shifts at the RIS are hereby configured as \vspace{-0.5em}
\begin{align}
	\label{eq:opt_IRS_closed_form_2}
	{{\phi}}_n^{\star}(k) = \left\{\begin{array}{ll}
		{2 \pi f_0 {\tilde{\tau}^{\mathrm{B}}_n}} , & \text { if } n \in \mathcal{N}_s(k), \\
		{2 \pi f_0 {\tilde{\tau}^{\mathrm{B}}_n}}  + \pi, & \text { otherwise}.
	\end{array}\right.
\end{align}
Thereby, { $\mathcal{N}_s (k)$ } is the {selected subset of reflective element indices with a size of $N_s$. Note that utilizing beamforming improves the strength of the received signal, while employing inverted beamforming degrades it. To ensure that Bob can always benefit from transmit beamforming and RIS, we specify $M_s > M - M_s$ and $N_s > N - N_s$, i.e., $M_s > \frac{M}{2}$ and $N_s > \frac{N}{2}$. By applying FDA and RIBES, we observe the following lemma.} 

\begin{lemma}
	\label{lemma:lemma1}
	The received SNR at Bob with {RIBES} is given by
	\begin{align}
		\label{eq:SNR_Bob_FDA_w_sel1}
		\gamma_{\mathrm{B}}^\star =  \frac{P}{\sigma_{\mathrm{B}}^2} L_\mathrm{G}(R_1) L_\mathrm{H}(R_{\mathrm{B}}) \frac{(2 M_s-M)^2}{M}(2N_s-N)^2.
	\end{align}
\end{lemma}
\begin{proof}
	The proof is provided in Appendix A.
\end{proof}

\begin{remark}
	{When comparing \eqref{eq:SNR_Bob_FDA} and \eqref{eq:SNR_Bob_FDA_w_sel1}, we observe that the received SNR at Bob is reduced if {RIBES} is employed. Nevertheless, the use of {RIBES} randomizes the received signal at Eve, significantly degrading the received SNR at Eve, {as we will see in the next section.} Consequently, the secrecy performance can be enhanced.}
\end{remark}
%{Details of using RIAS to improve the secrecy rate are introduced in the following section.}

%Here, we can see that the rece ived signal power is related to both $M_s$ and $N_s$. However, it remains constant over time.

\section{Secrecy analysis}
\label{sec:secrecy}
In this section, we first define the scaling factor of the received signal using FDA and {RIBES}, and then derive its statistical properties. Afterward, we examine the secrecy performance when employing FDA without {RIBES}. We demonstrate the advantages of utilizing FDA over traditional phased array while also highlighting the associated issues. Subsequently, we delve into an analysis of the secrecy rate when both FDA and {RIBES} are employed, and show how this approach resolves the issues in the utilization of FDA.

\subsection{Statistical properties  of the scaling factor}
It is noted that we work within a two dimensional geometric framework, where the positions of the BS and RIS remain fixed.  {As a result, the user's location can be specified using the parameter tuple $(R_{\text{UE}}, \theta_{\text{UE}})$, where $R_{\text{UE}}$ is the distance between the RIS and user, and  $\theta_{\text{UE}}$ is the AOA at the receiver, as introduced in Sec. \ref{sec:chmod}}. With FDA and {RIBES}, the received signal of Eve with $(R_{\mathrm{E}}, \theta_{\mathrm{E}})$ is
% e^{j 2 \pi f_0 \frac{R_{\text{Eve}}-R_{\text{Bob}}}{c}}
\begin{align}
	\label{eq:received_signal_Eve}
	y_{\mathrm{E}}^\star (k)&= 	\sqrt{P}{\mathbf{h}}_{\mathrm{E} }^\star(k)^H \boldsymbol{w}^{\star}(k) x(k) + n_{\mathrm{E}}(k) \nonumber \\
	& =    \sqrt{\frac{P L_\mathrm{G}(R_1) L_\mathrm{H}(R_{\mathrm{E}})}{M}}    u(k) v(k)    x(k) +n_{\mathrm{E}}(k),
\end{align}
where 
\begin{align}
%	\eta &=  \sum_{m \in \mathcal{M}_s}e^{j 2\pi 
%		f_0 \left( \tilde{\tau}_m^{d} - \tilde{\tau}_m \right) }  e^{j 2\pi 
%		\Delta	f_m \frac{R_d-R_{1}+R_{\text{Bob}}}{c}}  \nonumber \\ &- \sum_{m \notin \mathcal{M}_s} e^{j 2\pi 
%		f_0 \left( \tilde{\tau}_m^{d} - \tilde{\tau}_m \right) }  e^{j 2\pi 
%		\Delta	f_m \frac{R_d-R_{1}+R_{\text{Bob}}}{c}} , \\ 
	u(k) &= \!  \! \sum_{m \in \mathcal{M}_s(k)} \!e^{j 2\pi 
		\Delta	f_m \frac{R_{\mathrm{E}}-R_{\mathrm{B}}}{c}} \!\! - \!\! \sum_{m \notin \mathcal{M}_s(k)}\! e^{j 2\pi 
		\Delta	f_m  \frac{R_{\mathrm{E}}-R_{\mathrm{B}}}{c}} , \nonumber \\
	%\end{align}
	%\begin{align}
	v(k) &= 
	\sum_{n \in \mathcal{N}_s(k)} e^{j 2\pi 
		f_0 \left( \tilde{\tau}_n^{\mathrm{E}} - \tilde{\tau}_n^{\mathrm{B}} \right) } - \sum_{n \notin \mathcal{N}_s(k)} e^{j 2\pi 
		f_0 \left( \tilde{\tau}_n^{\mathrm{E}} - \tilde{\tau}_n^{\mathrm{B}} \right) }. \nonumber
\end{align}
{Due to the randomness of the selected subset $\mathcal{M}_s(k)$ and $\mathcal{N}_s(k)$, both $u(k)$ and $v(k)$ can be treated as random variables for Eve. It is important to note that for the legitimate receiver, Bob, $u(k) = 2M_s - M$ and $v(k) = 2N_s - N$ are constants throughout the transmission. Consequently, the randomness introduced by subset selection has no impact on Bob.  For the sake of readability and clarity, we omit the index $k$ in the remaining part of this work.}
% i.e., $u(k)$ and $v(k)$ are denoted as $u$ and $v$ in the following. 
 Let us define $\beta(R_{\mathrm{E}}, \theta_{\mathrm{E}})$ as a scaling factor of the received signal, and it is given by  
%e^{j 2 \pi f_0 \frac{R_{\text{Eve}}-R_{\text{Bob}}}{c}}
\begin{align}
	\beta  (R_{\mathrm{E}}, \theta_{\mathrm{E}}) =   \sqrt{\frac{L_\mathrm{G}(R_1) L_\mathrm{H}(R_{\mathrm{E}})}{M}}    u v ,
\end{align}
with which the received signal at Eve in \eqref{eq:received_signal_Eve} can be rewritten as \vspace{-1em}
\begin{align}
	\label{eq:received_signal_Eve2}
	y_{\mathrm{E}}^\star =\sqrt{P} \beta  (R_{\mathrm{E}}, \theta_{\mathrm{E}}) x +n_{\mathrm{E}}.
\end{align}
Here, we can see that $\beta  (R_{\mathrm{E}}, \theta_{\mathrm{E}}) $ is a complex-valued random variable.
\begin{remark}
	{Assuming we have $M = 21$ BS antennas and $N = 21$ reflective elements, for each symbol interval, we randomly select $M_s = 12$ antennas and $N_s = 12$ elements. This results in a total of $\binom{N}{N_s} \binom{M}{M_s} = 293,930^2$ possible combinations of $\beta $ at each transmit symbol. The extensive number of combinations makes it challenging for an eavesdropper to decode the transmitted symbol.}
	% Furthermore, assuming noiseless transmission and the use of QPSK modulation for the transmitted symbols, we can observe that the phase of the received constellations is randomized at the eavesdropper, whereas the phase of the received constellations is the same as the transmitted symbol, if the receiver noise is ignored.
	Therefore, the secrecy performance can be enhanced through the use of {RIBES}.
\end{remark}

%, even if the eavesdropper has significantly higher channel gain compared to Bob.
In the following lemma, we characterize the statistical properties of $\beta(R_{\mathrm{E}}, \theta_{\mathrm{E}})$ that are important for the analysis of secrecy performance. 
%
%	
%The core idea revolves around the random selection of a subset of transmit antennas and IRS elements to create a constructive signal combination. Simultaneously, the remaining antennas and IRS elements are strategically configured to create a destructive signal combination. The selection is randomized in each symbol interval. As a result, the desired signal received at the target receiver remains constant, while any other users receive a signal resembling noise, making it difficult for them to decode. 
%\begin{lemma}
%	\label{lemma2}
%	The mean value and the variance of the received signal gain $\beta(R_{\text{Eve}}, \theta_{\text{Eve}})$  with FDA, transmit antenna and IRS subset selection are, respectively, given by  
%	\begin{align}
	%		&\mathbb{E}[\beta  (R_{\text{Eve}}, \theta_{\text{Eve}})]  \nonumber \\
	%		&= \sqrt{\frac{P}{M}} \bigg(\sqrt{ \frac{ C_0}{ {R_d^{\alpha_d}}}} p(R_d) e^{j 2 \pi f_0 \frac{R_d-R_1-R_{\text{Bob}}}{c}} \frac{2M_s-M}{M} \mu_d \nonumber \\
	%		& + \frac{C_0}{\sqrt{ R_1^{\alpha_1}R_{\text{Eve}}^{\alpha_2}}}  e^{j 2 \pi f_0 \frac{R_{\text{Eve}}-R_{\text{Bob}}}{c}} \frac{\left( 2M_s-M\right) \left( 2N_s-N \right) }{MN}  \mu_1 \mu_2 \mu_3 \bigg) 
	%	\end{align}	
%and
%	\begin{align}
	%		&\mathbb{V}[\beta  (R_{\text{Eve}}, \theta_{\text{Eve}})]  \nonumber \\
	%		&= {\frac{P}{M}}\bigg( \frac{ C_0}{ {R_d^{\alpha_d}}} p(R_d)^2  4 \frac{M_s(M-Ms)}{M(M-1)}(M- \frac{1}{M} \mu_d^2) \nonumber \\
	%		& + \frac{C_0^2}{{ R_1^{\alpha_1}R_{\text{Eve}}^{\alpha_2}}}   \bigg) 
	%	\end{align}
%\end{lemma}
%\textcolor{red}{To do, check the distribution of beta}
\begin{lemma}
	\label{lemma:lemma2}
%{The scaling factor $\beta(R_{\mathrm{E}}, \theta_{\mathrm{E}})$, when employing FDA and RIAS combined, can be approximated as Gaussian. Thereby, the mean value of $\beta(R_{\mathrm{E}}, \theta_{\mathrm{E}})$} is
{The mean of the scaling factor $\beta(R_{\mathrm{E}}, \theta_{\mathrm{E}})$, when employing FDA and RIBES combined,}
%can be approximated as Gaussian. Thereby, the mean value of $\beta(R_{\mathrm{E}}, \theta_{\mathrm{E}})$} 
is  \vspace{-0.5em}
%	The mean value of the scaling factor $\beta(R_{\mathrm{E}}, \theta_{\mathrm{E}})$, when employing FDA and RIAS combined, is given by
	%  e^{j 2 \pi f_0 \frac{R_{\text{Eve}}-R_{\text{Bob}}}{c}} 
	\begin{align}
		\label{eq:mean_scaling_fator}
		&\mathbb{E}[\beta  (R_{\mathrm{E}}, \theta_{\mathrm{E}})]   \\&=  {\sqrt{\frac{L_\mathrm{G}(R_1) L_\mathrm{H}(R_{\mathrm{E}})}{M}}}
	\frac{\left( 2M_s-M\right) \left( 2N_s-N \right) }{MN}  \mu_1 \mu_2 \mu_3,\nonumber
	\end{align}	
    where \vspace{-1em}
    \begin{align}
    	\label{eq:mean_mu}
    	\mu_1 & = \frac{ \sin \left(M \pi\Delta F (R_{\mathrm{E}}-R_{\mathrm{B}}) /c\right)  }{\sin \left( \pi\Delta F (R_{\mathrm{E}}-R_{\mathrm{B}}) /c \right) },  \\
    \mu_2 & = \frac{ \sin \left( 0.5 N_\mathrm{H} \pi  ( \cos(\theta_{\mathrm{B}})-\cos(\theta_{\mathrm{E}}))\right)  }{\sin \left( 0.5 \pi   ( \cos(\theta_{\mathrm{B}})-\cos(\theta_{\mathrm{E}})) \right) },\\
    \mu_3 & = \frac{ \sin \left( 0.5 N_\mathrm{V} \pi  ( \sin(\theta_{\mathrm{E}})-\sin(\theta_{\mathrm{B}}))\right)  }{\sin \left( 0.5 \pi  ( \sin(\theta_{\mathrm{E}})-\sin(\theta_{\mathrm{B}})) \right) }.	
    \end{align}	
	The variance of $\beta(R_{\mathrm{E}}, \theta_{\mathrm{E}})$ is
	\begin{align}
		&\mathbb{V}[\beta  (R_{\mathrm{E}}, \theta_{\mathrm{E}})]  \\
		&=\frac{L_G(R_1) L_H(R_{\mathrm{E}})}{M}\big(	\mathbb{V}[u] \mathbb{V}[v] +	\mathbb{V}[v] \left| \mathbb{E}[u]\right| ^2 \! +\! \mathbb{V}[u] \left| \mathbb{E}[v]\right| ^2 \big)  ,  \nonumber
	\end{align}
	where \vspace{-0.5em}
	\begin{align}
		%	\mathbb{E}[\eta] &= \frac{2M_s-M}{M} \mu_d, 
		\mathbb{E}[u] &= \frac{2M_s-M}{M} \mu_1, 
		\quad	\mathbb{E}[v] = \frac{2N_s-N}{N} \mu_2 \mu_3, 	 \\
		%	\mathbb{V}[\eta] &= 4 \frac{M_s(M-Ms)}{M(M-1)}(M- \frac{1}{M} \mu_d^2), \\
		\label{eq:variance_u}
		\mathbb{V}[u] &= 4 \frac{M_s(M-Ms)}{M(M-1)}\left(M- \frac{1}{M} \mu_1^2\right), \\
		\label{eq:variance_v}
		\mathbb{V}[v] &= 4 \frac{N_s(N-Ns)}{N(N-1)}\left(N- \frac{1}{N} \mu_2^2 \mu_3^2\right).
	\end{align}
	%Also, when both $M$ and $N$ are sufficient large
\end{lemma}
\begin{proof}
	The proof is provided in Appendix B.
\end{proof}

\subsection{Secrecy rate with FDA and {RIBES}}
\begin{figure*}
	\centering
	\begin{subfigure}[b]{0.325\textwidth}
		\centering
		\includegraphics[width=1\linewidth]{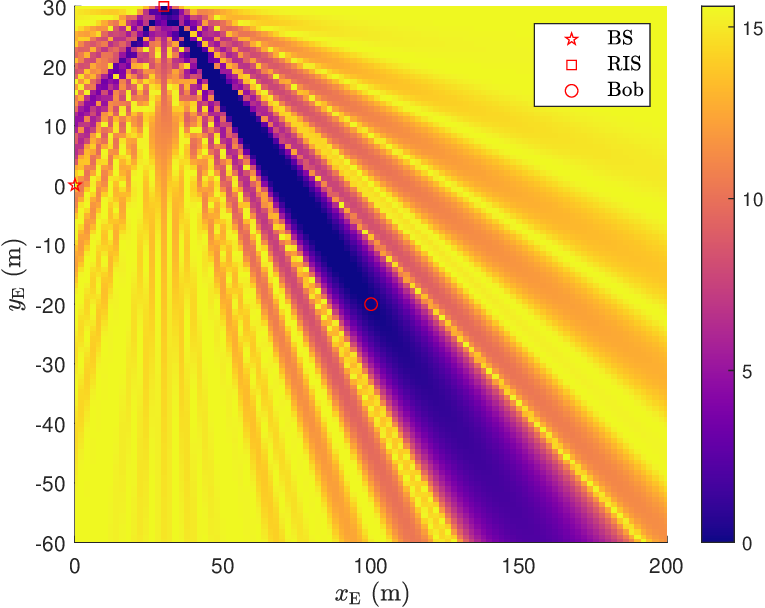}
		\caption{Conventional phased array}
		\label{fig:fig_sec_3_1} 
	\end{subfigure}
	\hfill
	\begin{subfigure}[b]{0.325\textwidth}
		\centering
		\includegraphics[width=1\linewidth]{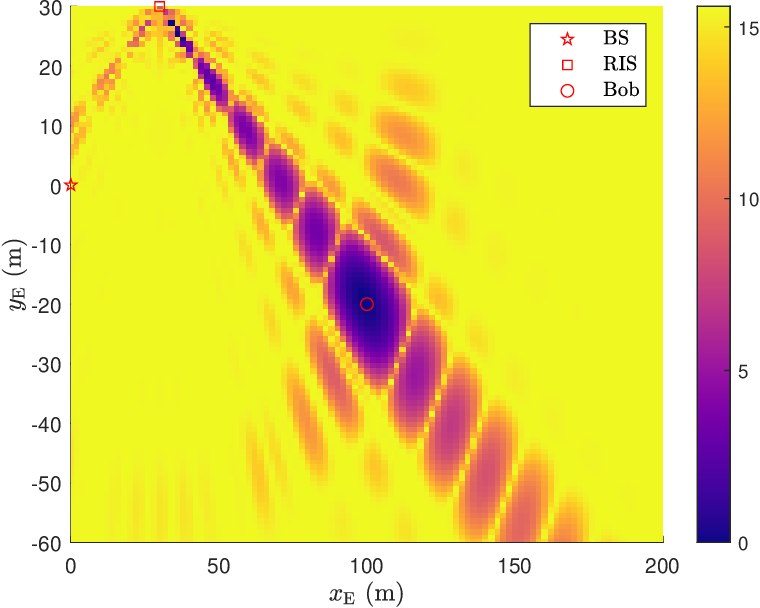}
		\caption{FDA with $\Delta F = 1$ MHz }
		\label{fig:fig_sec_3_2} 
	\end{subfigure} 
	\hfill
	\begin{subfigure}[b]{0.325\textwidth}
		\centering
		\includegraphics[width=1\linewidth]{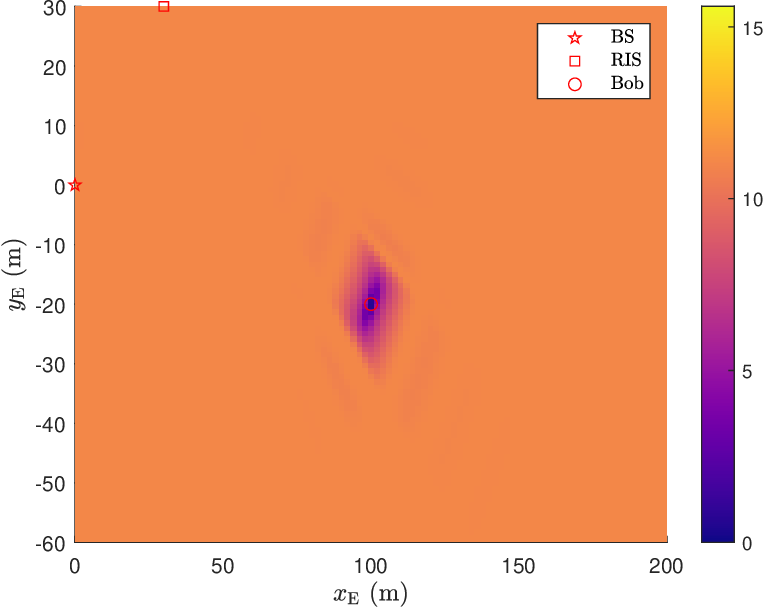}
		\caption{FDA with {RIBES}}
		\label{fig:fig_sec_3_3} 
	\end{subfigure} 
	\caption{{The heatmaps of the secrecy rate observed when Eve is  positioned at $[x_{\mathrm{E}}, y_{\mathrm{E}}]$ using various transmission techniques}}
	%with $M=17$, $N_H=N_V = 21$, $P = 5$ dBm, $\sigma_{\text{Bob}}^2 = \sigma_{\text{Eve}}^2 = -120$ dBm, $p(R_d) = 0$}
\label{fig:fig_sec_3} 
\vspace{-1em}
\end{figure*} 
Next, we investigate the secrecy rate with FDA but without {RIBES}.
\begin{proposition}
		For any eavesdropper {with the parameter tuple} $ \left( R_{\mathrm{E}}, \theta_{\mathrm{E}}\right) $,  the secrecy rate with FDA is given by
	\begin{align}
		\label{eq:secrecy_rate_1}
		 C_s (R_{\mathrm{E}}, \theta_{\mathrm{E}})&= \bigg[ \log_2 \left( 1 \!+\! \frac{P}{\sigma_{\mathrm{B}}^2} L_\mathrm{G}(R_1)L_\mathrm{H}(R_\mathrm{B})  M N^2 \right)  \nonumber  \\ & \hspace{-0.55cm} -\log_2 \left( 1\!+\!  \frac{P}{ \sigma_{\mathrm{E}}^2}   L_\mathrm{G}(R_1)L_\mathrm{H}(R_\mathrm{E})   \frac{\mu_1^2}{M}\mu_2^2\mu_3^2  \right) \bigg]^{+}.
	\end{align}		
\end{proposition}  
\begin{proof}
If {RIBES} is not utilized, the variance of the scaling factor $\beta(R_{\mathrm{E}}, \theta_{\mathrm{E}})$ becomes zero. Consequently, $\beta(R_{\mathrm{E}}, \theta_{\mathrm{E}})$ is a constant and can be calculated by employing \eqref{eq:mean_scaling_fator} while setting $M_s = M$ and $N_s = N$. Thus, the received SNR at Eve can be approximated as \vspace{-0.5em}
	\begin{align}
		\gamma_{\mathrm{E}} = \frac{P}{\sigma_{\mathrm{E}}^2} \left| \beta(R_{\mathrm{E}}, \theta_{\mathrm{E}}) \right|^2 =\frac{P}{ \sigma_{\mathrm{E}}^2}    L_\mathrm{G}(R_1)L_\mathrm{H}(R_\mathrm{E})  \frac{\mu_1^2}{M}\mu_2^2\mu_3^2. \nonumber	
	\end{align}	
 Furthermore, the received SNR at Bob without {RIBES}, i.e., $\gamma_{\mathrm{B}}$, has been approximated in \eqref{eq:SNR_Bob_FDA}. By substituting $\gamma_{\mathrm{B}}$ and $\gamma_{\mathrm{E}}$ into \eqref{eq:sec_cap}, we observe the secrecy rate without {RIBES}. This completes the proof. 
\end{proof}
We know that the maximum value of $\mu_1^2 \mu_2^2 \mu_3^2$ is $M^2 N_\mathrm{H}^2 N_\mathrm{V}^2$ achieved when  $(R_{\mathrm{E}}, \theta_{\mathrm{E}}) = (R_{\mathrm{B}}, \theta_{\mathrm{B}})$. Thus, the highest SNR at Eve is achieved when their position is identical to Bob.
%, which is impossible in practice. 
This observation signifies that, with FDA at the transmitter, we can ensure secure transmission within a range that takes both direction and distance into account. 
\begin{remark}
	When a conventional phased array is employed at the BS, the secrecy rate is given by
		\begin{align}
		\label{eq:secrecy_rate_conv}
		C_s^{conv} (\theta_{\mathrm{E}})&= \bigg[ \log_2 \left( 1 \!+\! \frac{P}{\sigma_{\mathrm{B}}^2} L_\mathrm{G}(R_1)L_\mathrm{H}(R_\mathrm{B})  M N^2 \right)   \nonumber \\ & \hspace{-0.5cm} -\log_2 \left( 1\!+\!  \frac{P}{ \sigma_{\mathrm{E}}^2}   L_\mathrm{G}(R_1)L_\mathrm{H}(R_\mathrm{E})   M\mu_2^2\mu_3^2  \right) \bigg]^{+}.
	\end{align}
This term holds because $\mu_1$ equals to $M$, if $\Delta F = 0$, according to \eqref{eq:mean_mu}. As a result, the secrecy rate with conventional phased array is only direction-dependent, {indicating that {secure} transmission is not possible if Eve is in the same direction as Bob and closer to the RIS.} 
\end{remark}
% However, when using a conventional phased array, $\mu_1$ is equal to $M$, indicating that the secrecy rate is only direction-dependent. Therefore, with a conventional phased array, {secure} transmission is possible only if Eve is not aligned with Bob. 
%\textcolor{blue}{ ("S"-shape, coupled phase and angle with uniform frequency shift )}
Here, we demonstrate that a direction-range secured communication can be achieved through the joint use of FDA, MRT transmit beamforming and RIS configuration, as proposed in Sec. \ref{sec:3_A}. However, there are still certain issues that need to be addressed.
%While employing the FDA, along with the utilization of the proposed transmit beamforming vector and IRS in Sec. \ref{sec:3_A}, provides security in terms of both angle and distance range, it does come with certain drawbacks. 
The scaling factor $\beta(R_{\mathrm{E}}, \theta_{\mathrm{E}})$ remains constant and increases linearly with the channel gain. An eavesdropper with a strong channel gain or very low receiver noise might have the potential to successfully decode the transmitted symbol. To tackle this issue, we apply {RIBES} in this work combined with FDA. 

We now analyze the secrecy rate when {RIBES} is applied. 
%Following \lemmaref{lemma:lemma2}, the received signal at Eve can be expressed as
%\begin{align}
%	&	y_{\mathrm{E}}^\star (R_{\mathrm{E}}, \theta_{\mathrm{E}}) = \sqrt{P}	\beta (R_{\mathrm{E}}, \theta_{\mathrm{E}})  x + n_{\mathrm{E}}, %\nonumber \\
%%	& = \sqrt{P}(\bar{\beta}(R_{\mathrm{E}}, \theta_{\mathrm{E}}) +\tilde {\beta}(R_{\mathrm{E}}, \theta_{\mathrm{E}}))x + n_{\mathrm{E}},
%\end{align}
%%where $\bar{\beta}(R_{\mathrm{E}}, \theta_{\mathrm{E}}) =\mathbb{E}[\beta  (R_{\mathrm{E}}, \theta_{\mathrm{E}})] $ and $\tilde {\beta}(R_{\mathrm{E}}, \theta_{\mathrm{E}})) =\beta (R_{\mathrm{E}}, \theta_{\mathrm{E}}) - \bar{\beta}(R_{\mathrm{E}}, \theta_{\mathrm{E}})  $ is a random variable with zero mean.
% Thus, the SNR observed at Eve can be computed as \cite{7876781}
%%\begin{align}
%%	\label{eq:SNR_Eve_w_random_Sel}
%%	\gamma_{\mathrm{E}}^\star &= \frac{P \left|\bar{\beta}(R_{\mathrm{E}}, \theta_{\mathrm{E}})  \right|^2 }{P\mathbb{V}[ \tilde{\beta}  (R_{\mathrm{E}}, \theta_{\mathrm{E}})] +\sigma_{\mathrm{E}}^2 } = \frac{P \left|\mathbb{E}[ {\beta}(R_{\mathrm{E}}, \theta_{\mathrm{E}})]  \right|^2 }{P\mathbb{V}[ {\beta}  (R_{\mathrm{E}}, \theta_{\mathrm{E}})] +\sigma_{\mathrm{E}}^2 }. 
%%\end{align}
{Following \eqref{eq:received_signal_Eve2} and applying \lemmaref{lemma:lemma2},} the SNR observed at Eve can be computed as \cite{7876781}
\begin{align}
	\label{eq:SNR_Eve_w_random_Sel}
	\gamma_{\mathrm{E}}^\star  = \frac{P \left|\mathbb{E}[ {\beta}(R_{\mathrm{E}}, \theta_{\mathrm{E}})]  \right|^2 }{P\mathbb{V}[ {\beta}  (R_{\mathrm{E}}, \theta_{\mathrm{E}})] +\sigma_{\mathrm{E}}^2 }. 
\end{align}
%
%Therefore, if an eavesdropper is situated in the same direction as Bob ($\theta_{\text{Bob}} = \theta_{\text{Eve}}$) and the eavesdropper has a higher channel gain than Bob ($R_{\text{Eve}} < R_{\text{Bob}}$), it becomes impossible to guarantee a positive secrecy rate. 
Thereby, the closed-form of the secrecy rate with FDA and {RIBES} can be observed by
\begin{align}
	C_s^\star(R_{\mathrm{E}}, \theta_{\mathrm{E}}) = \log_2(1+\gamma_{\mathrm{B}}^\star) -\log_2(1+\gamma_{\mathrm{E}}^\star),
\end{align}	
where $\gamma_{\mathrm{B}}^\star$ and $\gamma_{\mathrm{E}}^\star$ are stated in \eqref{eq:SNR_Bob_FDA_w_sel1} and \eqref{eq:SNR_Eve_w_random_Sel}, respectively. Note that when the channel gain of Eve {grows asymptotically high} or the receiver noise at Eve is sufficiently low, such that $\sigma_{\mathrm{E}}^2$ in \eqref{eq:SNR_Eve_w_random_Sel} can be ignored, the received SNR at Eve becomes
\begin{align}
	\gamma_{\mathrm{E}}^\star  \rightarrow  \frac{	\left( 2M_s-M\right)^2 \left( 2N_s-N \right)^2   \mu_1^2 \mu_2^2 \mu_3^2} {M^2 N^2 \big(	\mathbb{V}[u] \mathbb{V}[v] +	\mathbb{V}[v] \left| \mathbb{E}[u]\right| ^2 \! +\! \mathbb{V}[u] \left| \mathbb{E}[v]\right| ^2 \big)}.
\end{align}
This equation holds because the numerator and denominator of $\gamma_\mathrm{E}^\star$ in \eqref{eq:SNR_Eve_w_random_Sel} both linearly depend on the term of $L_\mathrm{G}(R_1) L_\mathrm{H}(R_{\mathrm{E}})$, according to \lemmaref{lemma:lemma2}. Consequently, the secrecy rate remains unaffected by the eavesdropper's channel gain. 
It is thus shown that {RIBES} is robust to high channel gains and low receiver noise at eavesdroppers. 

\subsubsection*{Example}
We show an illustration of the secrecy rate observed at coordinates  {$[x_{\mathrm{E}}, y_{\mathrm{E}}]$} when applying various transmission techniques in  \figref{fig:fig_sec_3}. The BS, RIS, and the target receiver Bob are located at [\SI{0}{\meter}, \SI{0}{\meter}], [\SI{30}{\meter}, \SI{30}{\meter}], and [\SI{100}{\meter}, \SI{-20}{\meter}], respectively. Furthermore, we set $M=21$, $N_\mathrm{H}=N_\mathrm{V} = 21$, $P = \SI{30}{dBm}$, $\sigma_{\mathrm{B}}^2 = \sigma_{\mathrm{E}}^2 = \SI{-120}{dBm}$.  \figref{fig:fig_sec_3_1} illustrates the secrecy rate when a conventional phased array is employed. We observe that the main beam is directed toward the target receiver Bob, and within the main beam region, the secrecy rate is very low. Notably, when Eve is positioned closer to the RIS than Bob, the secrecy rate within the main beam region drops to zero due to Eve's higher SNR compared to Bob. \figref{fig:fig_sec_3_2} shows the secrecy rate for the case that FDA are used with a fixed frequency increment of $\Delta F = \SI{1}{MHz}$.  
%\textcolor{blue}{ In this figure, a zero secrecy rate is observed if Eve is located nearby to Bob. Nevertheless, it is still possible to achieve a positive secrecy rate when Eve is located in the same direction as Bob and is closer to the RIS. However, we observe multiple local minima of the secrecy rate due to the high energy leakage on the slide lobes. Also, when Eve is close to the RIS, the secrecy rate drops to zero due to Eve's significantly higher channel gain compared to Bob.}
%\textcolor{blue}{ In this figure, we can see that it is possible to achieve a positive secrecy rate when Eve is located in the same direction as Bob and is closer to the RIS. However, we observe multiple local minima of the secrecy rate due to the high energy leakage on the slide lobes. Furthermore, when Eve is close to the RIS, the secrecy rate drops to zero due to Eve's significantly higher channel gain compared to Bob.}
{In this figure, we observe that a positive secrecy rate can be achieved when Eve is positioned in the same direction as Bob and is closer to the RIS. However, multiple local minima of the secrecy rate are present due to high energy leakage on the side lobes. Furthermore, when Eve is in close proximity to the RIS, the secrecy rate drops to zero due to Eve's significantly higher channel gain compared to Bob. These issues can be addressed by applying both FDA and RIBES, as shown in  \figref{fig:fig_sec_3_3}.   In this case, the overall secrecy performance is significantly enhanced compared to  \figref{fig:fig_sec_3_1} and  \figref{fig:fig_sec_3_2}. A global minimum of the secrecy rate is observed when Eve is at the {exact} same location as Bob. Furthermore, the secrecy rate observed for the case that Eve is in the proximity of Bob or RIS {is improved.} 
 %shows the secrecy rate when applying both FDA and RIAS.
%Fig. \ref{fig:fig_sec_3_3} shows the secrecy rate when applying both FDA and RIAS. We can see that the overall secrecy performance is significantly enhanced compared to Fig. \ref{fig:fig_sec_3_1} and Fig. \ref{fig:fig_sec_3_2}. A global minimum of the secrecy rate is observed when Eve located same as Bob. In addition, the secrecy rate observed in the proximity of Bob also improved.  
 However, the maximal secrecy rate is around \SI{10.5}{bits/sec/Hz} only, which is lower than the maximal secrecy rate of \SI{15}{bits/sec/Hz} observed in Fig. \ref{fig:fig_sec_3_1} and Fig. \ref{fig:fig_sec_3_2}. This is because the random subset selection randomizes the received signal at Eve, but also reduces the received signal power at Bob. In fact, there is a trade-off between the SNR at Bob and Eve with respect to  $M_s$ and $N_s$. Hence, $M_s$ and $N_s$ are optimized to maximize the secrecy rate in the following section.  

\section{Secrecy rate optimization}
\label{sec:optimization}
%\textcolor{blue}{Assuming no prior information about Eve's location is given, finding the optimal number of selected antennas $M_s$ and elements $N_s$ that maximize the secrecy rate is challenging. To address this challenge, we define the wiretap area for our setup in this section. Under assumption that}  
{If no prior information about Eve's location is given, it is challenging to find the optimal subset sizes of the transmit antennas and reflective elements that maximize the secrecy rate. To address this challenge, we define the wiretap area for our setup. Under the assumption that}  
 %To maximize the secrecy rate, we define the wiretap area for our setup in this section.
 the eavesdropper is located within the wiretap area, we derive the \emph{worst-case secrecy rate}, which is optimized by determining the optimal value of $M_s$ and $N_s$. 
%Then, we determine the optimal values for $M_s$ and $N_s$ to maximize the \textcolor{blue}{\textit{worst-case}} secrecy rate, assuming the eavesdropper is located within the wiretap area.
Furthermore, we optimize the frequency increment at the BS when the eavesdropper's location is given, and derive the upper bound for the secrecy rate. \vspace{-0.5em}
%To solve this problem, we find the lower bound of the secrecy rate and then determine the optimal $M_s$ and $N_s$ that maximize the lower bound. 
\subsection{Optimization without information of Eve's location}
%\textcolor{blue}{\st{Assuming no prior information about Eve's location is given, finding the optimal number of selected antennas $M_s$ and elements $N_s$ is challenging. To solve this problem, we find the \textit{worst-case} secrecy rate and then determine $M_s$ and $N_s$ that optimize the \textit{worst-case} secrecy rate. 
%To do this, }
{We now define the wiretap area for our setup, the boundary of which is determined by the {first null} point of the received SNR at Eve.}
From \lemmaref{lemma:lemma2} and \eqref{eq:SNR_Eve_w_random_Sel}, we can see that the null points of the received SNR at Eve with respect to $R_{\mathrm{E}}$ are determined by the null points of $\mu_1$, while the null points with respect to $\theta_{\mathrm{E}}$ is same as the null points of $\mu_2 \mu_3$. From \cite{kraus2002antennas}, the null points of $\mu_1$ occur when
% Given the fact that the null points of $\mu_1$ are \cite{kraus2002antennas}
\begin{align}
	\label{eq:null_points_mu_1}
	&2 \pi M \Delta F \frac{R_{\mathrm{E}}- R_{\mathrm{B}} }{c} = \pm 2 i \pi, \nonumber \\
	&\quad \text{s.t.}\ i \neq k M,\ i \in \mathbb{Z}^+, \ k \in \mathbb{Z}^+
\end{align} 
holds. We observe the first null of the received SNR at Eve is achieved when $\left| R_{\mathrm{E}}-R_{\mathrm{B}}\right| =  \Delta R =\frac{ c}{M \Delta F} $. Similarly, the null points of $\mu_2$ occur when 
\begin{align}
	 &N_\mathrm{H} \pi  ( \cos \theta_{\mathrm{B}}-\cos\theta_{\mathrm{E}})= \pm 2 i \pi,  \nonumber \\
	&\quad \text{s.t.}\ i \neq k N_\mathrm{H},\ i \in \mathbb{Z}^+, \ k \in \mathbb{Z}^+
\end{align} 
holds. Hence the first null of $\mu_2$ is obtained when  $\theta_{\mathrm{E}} = \arccos(\cos\theta_{\mathrm{B}} \pm \frac{2}{N_\mathrm{H}})$, indicating that we have 
\begin{align}\left| \theta_{\mathrm{E}}-\theta_{\mathrm{B}}\right| = 	\Delta \theta_1= \left| \arccos \left( \cos \theta_{\mathrm{B}} \pm \frac{2}{N_\mathrm{H}}\right)  -\theta_{\mathrm{B}}\right|. \nonumber
\end{align} 
 Similarly, the first null of $\mu_3$ is obtained when 
\begin{align}\left| \theta_{\mathrm{E}}-\theta_{\mathrm{B}}\right| = 	\Delta \theta_2 = \left| \arcsin \left( \sin\theta_{\mathrm{B}} \pm \frac{2}{N_\mathrm{V}}\right)  -\theta_{\mathrm{B}}\right|.\nonumber
 \end{align}
  As the received SNR at Eve is determined by the product of $\mu_2$ and $\mu_3$, the first null of Eve's received SNR concerning $\theta_{\mathrm{E}}$ is determined by $\min\left( \Delta \theta_1, \Delta \theta_2\right)$. Thus, the wiretap area and the corresponding target area are defined as follows.
\begin{definition}
	\label{definition:wiretap}
	The wiretap area in an RIS-assisted mmWave/THz transmission with FDA is 
	\begin{align}
		\label{eq:wiretap_area}
		\mathcal{R} = &\Big\lbrace (R_{\mathrm{E}}, \theta_{\mathrm{E}})   \Big| \left| R_{\mathrm{E}}-R_{\mathrm{B}}\right| \geqslant  \Delta R, \nonumber \\ &  \qquad  \qquad \quad \left| \theta_{\mathrm{E}}-\theta_{\mathrm{B}}\right| \geqslant  \min ( \Delta \theta_1 ,  \Delta \theta_2)
		\Big\rbrace, 
	\end{align}
	where $ \Delta R$, $\Delta \theta_1 $ and  $\Delta \theta_2 $ are, respectively, given by
	\begin{align}
		\Delta R &= \frac{ c}{M \Delta F},	\nonumber \\
		\Delta \theta_1& = \left| \arccos \left( \cos \theta_{\mathrm{B}} \pm \frac{2}{N_\mathrm{H}}\right)  -\theta_{\mathrm{B}}\right|, \nonumber \\
		\Delta \theta_2& = \left| \arcsin \left( \sin \theta_{\mathrm{B}} \pm \frac{2}{N_\mathrm{V}}\right)  -\theta_{\mathrm{B}}\right|. 
	\end{align}
\end{definition}
\begin{definition}
\label{definition:target}
%	\textcolor{orange}{The complementary set of the wiretap area, i.e., $\mathcal{R}$, defines the target area in an IRS-assisted mmWave/THz transmission with FDA. }
{The complementary set of the wiretap area $\mathcal{R}$ defines the target area.} 	
\end{definition}

\begin{remark}   
%the wiretap area remains the same with and without	the use of RIAS. This is because, the null points of Eve's received SNR are determined by the parameters $\mu_1$, $\mu_2$, and $\mu_3$ in both cases. Furthermore, 
The size of the target area decreases with the increasing number of transmit antennas and reflective elements, indicating a more focused transmission.
% Within the target area, the received SNR at Eve is strictly \textcolor{orange}{concave}, reaching its maximum when Eve is in the same position as Bob. 
Within the target area, the received SNR at Eve reaches its maximum when Eve is in the same position as Bob, and decreases with an increasing distance.
Furthermore, in the target area, the channel gains of Bob and Eve are nearly identical. Therefore, it is reasonable to assume that a {secure} transmission can always be achieved if the eavesdropper is in the wiretap area. 
\end{remark}
In this work, we assume that the eavesdropper is located in the wiretap area. Under this assumption, the \textit{worst-case secrecy rate} can be expressed as
%\begin{align}
%	\label{eq:lower_bound_secrecy}
%    C_s^{\text{LB}} & = \min_{(\theta_{\text{Eve}}, R_{\text{Eve}}) \in	\mathcal{R}} \left( \log_2(1+\gamma_{B}^r) -  \log_2(1+\gamma_{E}^r)\right)  \nonumber \\
%    & =\log_2(1+\gamma_{B}^r) - \max_{(\theta_{\text{Eve}}, R_{\text{Eve}}) \in	\mathcal{R}} \log_2(1+\gamma_{E}^r),
%\end{align}
\begin{align}
	\label{eq:lower_bound_secrecy}
	C_s^{\text{W-C}} & 
%	= \min_{(\theta_{{E}}, R_{{E}}) \in	\mathcal{R}} \left( \log_2(1+\gamma_{B}^r) -  \log_2(1+\gamma_{E}^r)\right)  \nonumber \\
	& =\log_2(1+\gamma_\mathrm{B}^\star) - \max_{(\theta_{\mathrm{E}}, R_{\mathrm{E}}) \in	\mathcal{R}} \log_2(1+\gamma_{\mathrm{E}}^\star). 
\end{align}
%where the second equality holds because $\gamma_{B}^r$ is unrelated to the location of Eve.
It is noticed that there is a trade-off between the first part and the second part of \eqref{eq:lower_bound_secrecy} when RIBES is applied. Given that we assume $M_s > \frac{M}{2}$ and $N_s > \frac{N}{2}$, when $M_s$ and $N_s$ increase, the rate at Bob, i.e., $\log_2(1+\gamma_\mathrm{B}^\star)$, also increases, while the negative rate at Eve, i.e., $-\log_2(1+\gamma_\mathrm{E}^\star)$, decreases. The reason is that the randomness of the subset selection is maximized when exactly half of the transmit antennas and half of the reflective elements are chosen, as the maximal value of $\binom{N}{N_s} \binom{M}{M_s}$ is achieved when $N_s = \frac{N}{2}$ and $M_s = \frac{M}{2}$. 
%Hence, maximal variance of $u$ and $v$ in \eqref{eq:variance_u} and \eqref{eq:variance_v} is achieved when $N_s = \frac{N}{2}$ and $M_s = \frac{M}{2}$. 
As a result, $-\log_2(1+\gamma_\mathrm{E}^\star)$ achieves its maximum when $N_s = \frac{N}{2}$ and $M_s = \frac{M}{2}$. Therefore, in the following we aim to determine the optimal values of $M_s$ and $N_s$ that maximize the \textit{worst-case secrecy rate}. Hence, the problem of interest is formulated as 
%\begin{align}
%	\label{eq:max_min_P2}
%	\text{P:} \quad	\max_{M_s, N_s}   \quad \left(  \log_2(1+\gamma_{B}^r) - \max_{(\theta_{\text{Eve}}, R_{\text{Eve}}) \in	\mathcal{R}} \log_2(1+\gamma_{E}^r)\right).
%\end{align}
\begin{align}
	\label{eq:max_min_P2}
 \quad	\max_{M_s, N_s}   \quad  C_s^{\text{W-C}} (M_s, N_s) .
\end{align}
To solve this problem, we now derive the upper bound of the received SNR at Eve.
% \textcolor{blue}{Analysis of the  trade-off ?}
%In the following, we aim to find a lower bound of the secrecy rate when the eavesdropper is located in the wiretap area. 
%
%an upper bound for the received SNR at Eve when he is located in the wiretap area. The problem of interest is now formulated as 
%\begin{align}
%	\label{eq:max_min_P2}
%\text{P:} \quad	\max_{M_s, N_s}   \quad \left(  \log_2(1+\gamma_{B}^r) - \max_{(\theta_{\text{Eve}}, R_{\text{Eve}}) \in	\mathcal{R}} \log_2(1+\gamma_{E}^r)\right).
%\end{align}
\begin{lemma}
	\label{lemma:upper_bound}	
	With FDA and RIBES, and for any eavesdropper within the wiretap area, we observe
%	 \begin{align}
%	 	\label{eq:UB1}
%	\gamma_{\mathrm{E}}^\star \leqslant \gamma_{E}^{\text{UB},1} =  \frac{(2M_s-M)^2(M-1)}{4M^2 M_s(M-M_s)\sin^2(\frac{3 \pi}{2 M})},  \theta_{\mathrm{E}}=\theta_{\mathrm{B}},
%\end{align} 
%\begin{align}
%	\label{eq:UB2}
%	&\gamma_{\mathrm{E}}^\star \leqslant \gamma_{E}^{\text{UB},2} =  \frac{(2N_s-N)^2(N-1){\lambda^2}}{4N^2 N_s(N-N_s)},  R_{\mathrm{E}}= R_{\mathrm{B}},
%\end{align}
\begin{align}
	\label{eq:UB1}
	&\gamma_{\mathrm{E}}^\star \leqslant \gamma_{E}^{\mathrm{UB},1} \! = \! \frac{(2M_s-M)^2(M-1)}{4 M_s(M-M_s) {(M^2\sin^2(\frac{3 \pi}{2 M})-1)}},  \theta_{\mathrm{E}}=\theta_{\mathrm{B}}, \\
%\end{align} 
% \begin{align}
 	\label{eq:UB2}
 	&\gamma_{\mathrm{E}}^\star \leqslant \gamma_{E}^{\mathrm{UB},2} =  \frac{(2N_s-N)^2(N-1)\lambda^{2}}{4 N_s(N-N_s) {(N^2 -\lambda^{2})}},  R_{\mathrm{E}}= R_{\mathrm{B}},
 \end{align}
{where $\lambda$ is the maximum of $\mu_2 \mu_3$ in the wiretap area, and it can be approximated by} \vspace{-0.5em}
% \begin{align}
% 	\label{eq:appro_lambda}
%&\lambda =  \max \\ & \Bigg( \frac{ \sin \left( 0.5 N_\mathrm{V} \pi  ( \sqrt{1-(\cos{\theta_{\mathrm{B}}} \pm \frac{3}{N_\mathrm{H}})^2}-\sin(\theta_{\mathrm{B}}))\right)  }{\sin(\frac{3 \pi}{2 N_\mathrm{H}})\sin \left( 0.5 \pi  ( \sqrt{1-(\cos{\theta_{\mathrm{B}}} \pm \frac{3}{N_\mathrm{H}})^2}-\sin(\theta_{\mathrm{B}}))\right) } , \nonumber \\ &
%\frac{ \sin \left( 0.5 N_\mathrm{H} \pi  ( \cos(\theta_{\mathrm{B}})-\sqrt{1-(\sin{\theta_{\mathrm{B}}} \pm \frac{3}{N_\mathrm{V}})^2})\right)  }{ \sin(\frac{3 \pi}{2 N_\mathrm{V}}) \sin \left( 0.5  \pi  ( \cos(\theta_{\mathrm{B}})-\sqrt{1-(\sin{\theta_{\mathrm{B}}} \pm \frac{3}{N_\mathrm{V}})^2})\right)  }
%\Bigg). \nonumber	
% \end{align}
% \begin{align}
%	\label{eq:appro_lambda}
%	\lambda =  \max  \Bigg( \frac{ \sin \left(  N_\mathrm{V} \Lambda_1 \right)  }{\sin(\frac{3 \pi}{2 N_\mathrm{H}})\sin \left(  \Lambda _1\right)  } , 
%	\frac{ \sin \left(  N_\mathrm{H}  \Lambda_2 \right)  }{ \sin(\frac{3 \pi}{2 N_\mathrm{V}}) \sin \left(\Lambda_2\right)  }
%	\Bigg). \nonumber	
%\end{align}
 \begin{align}
 	\label{eq:appro_lambda}
 	\lambda {\approx} \max\left( \frac{\sin\left( N_{\mathrm{V}} \Lambda_\mathrm{H} \right) }{\sin\left( \frac{3 \pi}{2 N_\mathrm{H}}\right)  \sin\left( \Lambda_\mathrm{H}\right) },  \frac{\sin\left( N_{\mathrm{H}} \Lambda_\mathrm{V} \right) }{\sin\left( \frac{3 \pi}{2 N_\mathrm{V}}\right)  \sin\left( \Lambda_\mathrm{V}\right) } \right), 	 
\end{align}	\vspace{-1em}
where  \vspace{-0.5em}
\begin{align}
\Lambda_\mathrm{H} &= 	 0.5 \pi  \left( \sqrt{1-\left(\cos{\theta_{\mathrm{B}}} \pm \frac{3}{N_\mathrm{H}}\right)^2}-\sin\theta_{\mathrm{B}}\right),\ \nonumber \\
\Lambda_\mathrm{V} &= 0.5 \pi  \left( \cos\theta_{\mathrm{B}}-\sqrt{1-\left(\sin{\theta_{\mathrm{B}}} \pm \frac{3}{N_\mathrm{V}}\right)^2} \right). \nonumber
\end{align}	
\end{lemma}
\begin{proof}
	The proof is provided in Appendix C.
\end{proof}
From \lemmaref{lemma:upper_bound}, we can see that if $ \theta_{\mathrm{E}}=\theta_{\mathrm{B}}$, the upper bound depends only on $M$ and $M_s$. Also, if $R_{\mathrm{E}}= R_{\mathrm{B}}$, the upper bound is only related to $N$ and $N_s$. Therefore, we divide the problem in \eqref{eq:max_min_P2} into two sub-problems, expressed as 
 \begin{align}
	\label{eq:max_min_P3}
	&\max_{M_s}   \quad   \log_2\left( 1+\gamma_{\mathrm{B}}^\star\right) - \log_2\left( 1+ \gamma_{\mathrm{E}}^{\text{UB},1} \right), \\
%\end{align}
%\begin{align}
	\label{eq:max_min_P4}
	&\max_{N_s}   \quad   \log_2\left( 1+\gamma_{\mathrm{B}}^\star\right) - \log_2\left( 1+ \gamma_{\mathrm{E}}^{\text{UB},2} \right),
\end{align}
respectively. { Here, the objective of \eqref{eq:max_min_P3} can be considered as a function of $M_s$ given $N_s$. Thus, the optimal value for $M_s$ can be found through the zero of the first derivative of the objective. The solution to \eqref{eq:max_min_P4} can be obtained using a similar approach.  Hence, we observe the following theorem. }
\begin{theorem}
	\label{theorem:opt_M_s_N_s}
	In RIS-assisted secure transmission with FDA and {RIBES}, {the optimal subset size of the selected antennas} to maximize the \textit{worst-case secrecy rate} is
	 \begin{align}
	 	\label{eq:opt_M_s}
	 	M_s^* &\!=\!  \left\lfloor\frac{M}{2} \left(1\!+ \!\sqrt{\frac{1\!-\!\frac{1}{M} \sqrt{\eta _\mathrm{B}^{-1}\eta_\mathrm{E}(\eta _\mathrm{E}\!-1\!)\!+\!M^2}}{1-\eta_\mathrm{E}}} \right) \right\rceil, 
%	 	\nonumber  \\ 
%	 	& \approx \left \lfloor   \frac{M}{2} \left(1+ \frac{1}{\sqrt{1+\sqrt{\beta_E}}} \right) \right \rfloor ,
    \end{align}
where $\eta_\mathrm{B} = \frac{P}{M\sigma_{\mathrm{B}}^2} L_\mathrm{G}(R_1)L_\mathrm{H}(R_\mathrm{B}) (2N_s-N)^2 $ and $\eta_\mathrm{E} = \frac{(M-1)}{M^2 \sin^2(\frac{3 \pi}{2 M}){-1}}$. Furthermore, {the optimal subset size of the selected elements} is given by
	 \begin{align}
	 		\label{eq:opt_N_s}
	N_s^* &\!= \!   \left\lfloor\frac{N}{2} \left(1 \! + \!\sqrt{\frac{1\!-\!\frac{1}{N} \sqrt{{\zeta_\mathrm{B}}^{-1}{\zeta_\mathrm{E}}({\zeta_\mathrm{E}}\!-1\!)\!\!+N^2}}{1-{\zeta_\mathrm{E}}}} \right)  \right\rceil,
%	 \nonumber  \\ 
%	& \approx \left \lfloor   \frac{N}{2} \left(1+ \frac{1}{\sqrt{1+\sqrt{\tilde{\beta_E}}}} \right) \right \rfloor ,
\end{align}
where  ${\zeta}_\mathrm{B} = \frac{P}{M\sigma_{\text{B}}^2} L_\mathrm{G}(R_1)L_\mathrm{H}(R_\mathrm{B})  (2M_s-M)^2 $ and %$\tilde{\beta}_\mathrm{E} = \frac{(N-1)\lambda^2}{N^2}$.
${\zeta}_\mathrm{E} = \frac{(N-1)\lambda^2}{{N^2-\lambda^2}}$.
\end{theorem}
\begin{proof}
	The proof is provided in Appendix D.
\end{proof}
%From Theorem \ref{theorem:opt_M_s_N_s}, we can determine the optimal $M_s$ given $N_s$, wheres the optimal $N_s$ can be determined given $M_s$. Thus, we   
\begin{proposition}
	\label{proposition:opt_M_s_N_s}
When $\gamma_\mathrm{B}^\star$ is significantly larger than one, i.e., \SI{0}{dB}, {Theorem} \ref{theorem:opt_M_s_N_s} can be simplified as
	 \begin{align}
	 \label{eq:optimal_M_s_appr}
  	M_s^* &=    \left\lfloor  {\frac{M}{2} \left(1+ \frac{1}{\sqrt{1+\sqrt{\eta_\mathrm{E}}}} \right)}  \right\rceil , \\
%   \end{align}
%	 \begin{align}
	\label{eq:optimal_N_s_appr}
	N_s^* &=   \left\lfloor  \frac{N}{2} \left(1+ \frac{1}{\sqrt{1+\sqrt{{\zeta_\mathrm{E}}}}} \right) \right\rceil .
\end{align}
\begin{proof}
%	By using $1+\gamma_{B}^r \approx \gamma_{B}^r $, we can observe \eqref{eq:optimal_M_s_appr} and \eqref{eq:optimal_N_s_appr} following the same way as used in the proof of Theorem \ref{theorem:opt_M_s_N_s}. Thus, we omit the proof due to space constraints.
	The proposition can be proven similar to the proof provided for Theorem \ref{theorem:opt_M_s_N_s} in Appendix C. Thereby, the approximation $1+\gamma_\mathrm{B}^\star \approx \gamma_\mathrm{B}^\star $  is used.
\end{proof}
\end{proposition}
From {Proposition} \ref{proposition:opt_M_s_N_s},  we can easily determine {the optimal subset size of the selected antennas} based on $M$ for the case that $\gamma_\mathrm{B}^\star\gg \SI{0}{dB}$. For the same case, {the optimal subset size of the selected elements} is only dependent on $N$.  {It is worth noting that $\gamma_\mathrm{B}^\star \gg \SI{0}{dB}$ holds true in RIS-assisted systems with a sufficiently large number of elements \cite{9343768, 9091552, 9743437}.}
\subsection{Optimization given Eve's Location}
The secrecy rate can be further enhanced when Eve's location is known at the BS. Assuming that the frequency increments between neighboring antennas can be perfectly adjusted, we can find an optimal frequency increment such that the received SNR of Eve is {always $-\infty$ \SI{}{dB}}. 
%The null points of $\mu_1$ are given by 
%
%\begin{align}
%	&2 \pi M \Delta F \frac{R_{\text{Eve}}- R_{\text{Bob}} }{c} = \pm 2 i \pi, \nonumber \\
%	&\quad \text{s.t.}\ i \neq k M,\ i \in \mathbb{Z}^+, \ k \in \mathbb{Z}^+.
%\end{align}
%we can design the optimal frequency increment that null the received SNR at Eve
According to \eqref{eq:null_points_mu_1}, the mean value of $\beta(R_{\mathrm{E}}, \theta_{\mathrm{E}})$ becomes zero, if 
\begin{align}
	\label{eq: optimal_frequency_increment}
	\Delta F^* &= \frac{\pm i c}{M (R_{\mathrm{E}}- R_{\mathrm{B}})}, \quad \frac{M-1}{2} \Delta F^* \leqslant 10^{-3} f_0 .
\end{align}
Consequently, the received SNR at Eve is {$-\infty$ \SI{}{dB}}. 
In this context, the achievable secrecy rate is equivalent to the communication rate at Bob, i.e.,
%\begin{align}
%	\label{eq:sec_rate_no_hd_opt}
%	C_s^{*} =& \log_2 \left( 1+ \frac{P}{\sigma_{\text{Bob}}^2} \frac{C_0^2}{{R_1^{\alpha_1}R_{\text{Bob}}^{\alpha_2}}} M N^2 \right).
%\end{align}
\begin{align}
	\label{eq:sec_rate_no_hd_opt2}
	C_s^{\star} =  \log_2 (1+\gamma_\mathrm{B}^\star) &=\log_2\bigg( 1+ \frac{P}{\sigma_{\mathrm{B}}^2} L_\mathrm{G}(R_1)L_\mathrm{H}(R_\mathrm{B}) \nonumber  \\
	& \quad {\times \frac{(2 M_s-M)^2}{M}(2N_s-N)^2 \bigg)}. 
\end{align}
Note that, for the case that Eve's location is known and the optimally designed frequency increment is employed, the maximal secrecy rate will be achieved if $M_s = M$ and $N_s = N$, indicating that {RIBES} does not need to be employed. Then, the secrecy rate becomes
\begin{align}
	\label{eq:sec_rate_UB}
	C_s^{\text{UB}} =& \log_2 \left( 1+ \frac{P}{\sigma_{\mathrm{B}}^2} L_\mathrm{G}(R_1) L_\mathrm{H}(R_\mathrm{B}) M N^2 \right).
\end{align}
%optimal $M_s$ and $N_s$ to maximize the secrecy rate is $M$ and $N$, respectively. 
%By comparing \eqref{eq:sec_rate_no_hd_opt} and \eqref{eq:sec_rate_no_hd_opt2},  it is evident that when the frequency increment can be precisely controlled, the secrecy rate is higher when random subset selection is not employed compared to when it is used. 
{Note that, achieving a perfect frequency increment is not always possible due to hardware limitations and the constraint of the maximum frequency increment, i.e., $\max (\Delta F_m) < 10^{-3} f_0$. Nevertheless, \eqref{eq:sec_rate_UB} is still an upper bound of the secrecy rate.}

\section{Numerical results}
\label{sec:numerical_results}
In this section, we present the numerical results associated with the analytical results derived in this work. Throughout the simulations, we set $P = \SI{30}{dBm}$, $\sigma_{\mathrm{B}}^2 = \sigma_{\mathrm{E}}^2 = \SI{-120}{dBm}$. {The path loss of the mmWave/THz channel is modeled as
$
	   %	L (R) = 60 + 20 \log_{10}(R) [\mathrm{dB}],
	   	L(R) = L_0 + 10 \alpha \log_{10}(R) \ [\mathrm{dB}],
$	
where $R$, $L_0$ and $\alpha$ are the propagation distance, path loss value at a reference distance of \SI{1}{\meter}, and path loss exponent. Here, we employ a large reference path loss $L_0 = \SI{60}{dB}$ to align with the propagation conditions of mmWave/THz. Additionally, we set $\alpha$ to 2.} The center frequency is $f_0 = 60$ GHz. Moreover, the BS, RIS and Bob are located at [\SI{0}{\meter}, \SI{0}{\meter}], [\SI{30}{\meter} \SI{30}{\meter}], and [\SI{100}{\meter}, \SI{-20}{\meter}], respectively. From this parameter choice follows that $\theta_{\mathrm{B}} = \SI{0.62}{rad}$ and $R_\mathrm{B} = \SI{86.02}{\meter}$. 

\begin{figure}
	\centering
	\includegraphics[width=0.8\linewidth]{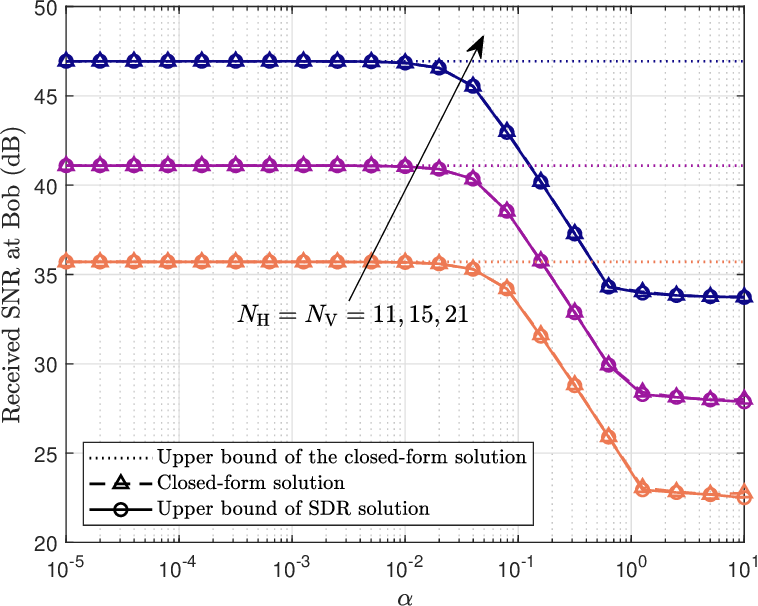}
	\caption{{Comparison of received SNR at Bob using different RIS solutions with various number of reflective elements}}
	\label{fig:evaluationofclosedformirs} \vspace{-1em}
\end{figure}
We first evaluate the performance of the proposed closed-form solution of $\phi_n^{\mathrm{opt}}$ stated in \eqref{eq:opt_IRS_closed_form} in terms of the received SNR at Bob. In \figref{fig:evaluationofclosedformirs}, we compare the {accurate simulation results using the closed-form solution with the upper bound in \eqref{eq:SNR_Bob_FDA}}, and the upper bound obtained by solving the SDR problem in \eqref{eq:sdp}. The SNR is presented over $\alpha = \frac{\max(\Delta F_m)}{f_0}$, i.e., the ratio of the maximal frequency shift and the center frequency. From this figure, we observe that the results obtained using the closed-form solution almost align with the upper bound achieved through SDR. 
%\textcolor{blue}{Only when the ratio is very large, the approach based on SDR slightly outperforms that based on the closed-form solution.} 
Additionally, {the performance gap between the upper bound of the closed-form solution and the accurate simulation result is negligible and can be disregarded when $\alpha$ is less than $10^{-2}$.}
{Hence, in the following results, we set $\max(\Delta F_m) < 10^{-3} f_0$.}

\begin{figure}
	\centering
	\includegraphics[width=0.8\linewidth]{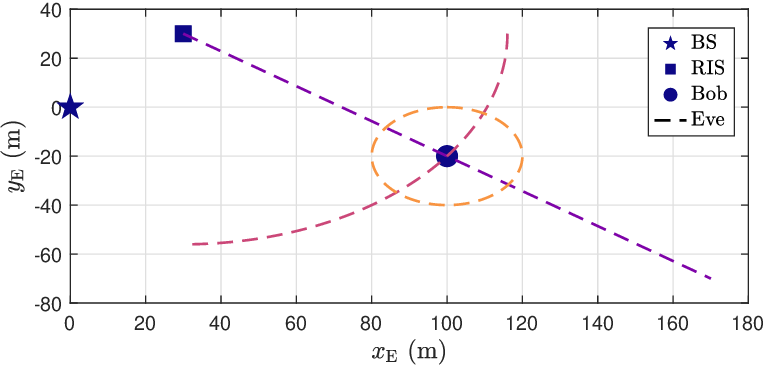}
	\caption{Considered locations of all devices. Dashed lines: potential locations of Eve}
	\label{fig:trajplot} \vspace{-1.5em}
\end{figure}
In the following subsections, we investigate the received SNRs at Bob and Eve and the resulting secrecy rate under the assumption that Eve is located along different, beneficial paths, as depicted in \figref{fig:trajplot}. In more details, first, the case where the AOA at Bob is the same as the one of Eve, i.e., $\theta_{\mathrm{B}} = \theta_{\mathrm{E}}$, is investigated (purple path). Second, Bob and Eve are assumed to be located equidistant from the RIS, i.e., $R_{\mathrm{B}} = R_{\mathrm{E}}$, (red path). Finally, a scenario is considered where Eve is located on a circle centered at Bob (orange path). In these scenarios, we set $M = N_\mathrm{H} = N_\mathrm{V} = 21$ and $\Delta F = \SI{1}{MHz}$ (if not specified otherwise). 
\vspace{-1em}
\subsection{{Equal AOA at Bob and Eve} }
\begin{figure}
	\centering
	\includegraphics[width=0.8\linewidth]{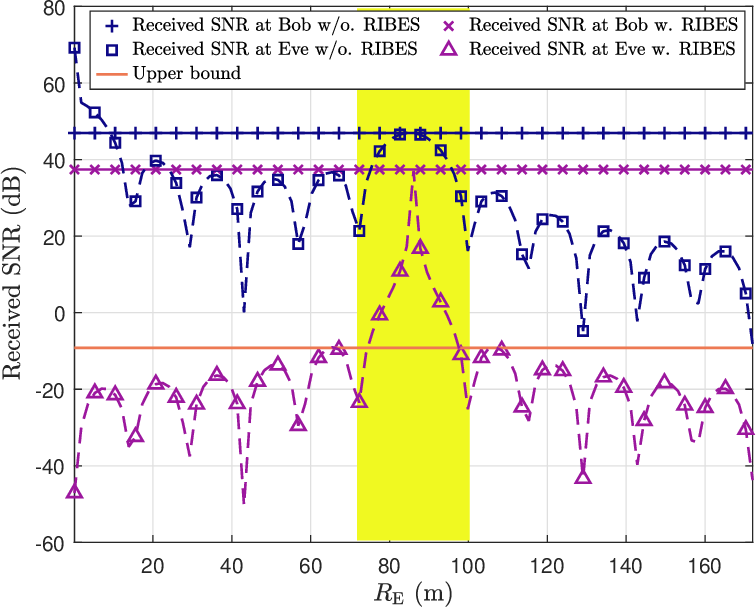}
	\caption{Comparison of the received SNR at Bob and Eve with different transmission techniques,  $M_s = \left \lfloor \frac{2}{3} M \right \rceil $. Markers: simulation results;
		curves: theoretical results.}
	\label{fig:receivedsnrvsub1} \vspace{-1.5em}
\end{figure}
We now examine the performance of the proposed transmission technique when $\theta_{\mathrm{B}} = \theta_{\mathrm{E}}$. In this case, according to Lemma \ref{lemma:lemma2}, random elements subset selection at the RIS does not provide any benefit. Thus, we set $N_s = N$ (if not specified otherwise). 
%The yellow-shaded area in the middle of the figure (the vertical strip between ?? m and 100 m)

The received SNR at Bob and Eve with different transmission techniques with respect to $R_{\mathrm{E}}$ is illustrated in \figref{fig:receivedsnrvsub1}. {The yellow-shaded area in the middle of the figure (the vertical strip between \SI{71.74}{\meter} and \SI{100.31}{\meter})} shows {the target area, according to \defref{definition:target}.} The perfect match between the curves and markers in this figure indicates the correctness of the closed-form expressions. Further, we observe that the received SNR at Bob is slightly reduced when {RIBES} is applied, while the received SNR at Eve is significantly degraded, especially in the wiretap area.  
In addition, as $R_{\mathrm{E}}$ decreases, the peaks of the received SNR at Eve increase for the case that {RIBES} is not employed. In particular, without {RIBES} and when $R_{\mathrm{\mathrm{E}}}$ is less than \SI{10}{\meter}, the received SNR at Eve exceeds that at Bob. This is because the channel gain of Eve is significantly larger than that of Bob. However, when {RIBES} is applied, we observe that the values of the peaks decrease with decreasing $R_{\mathrm{E}}$. As a result, Eve always has a lower SNR than Bob, indicating secure transmission can be achieved.  We also plot the upper bound computed by  \eqref{eq:UB1} in Lemma \ref{lemma:upper_bound}, which agrees well with the maximum received SNR at Eve in the wiretap area when {RIBES} is applied.
\begin{figure}
	\centering
	\begin{subfigure}[b]{0.5\textwidth}
		\centering
		\includegraphics[width=0.8\linewidth]{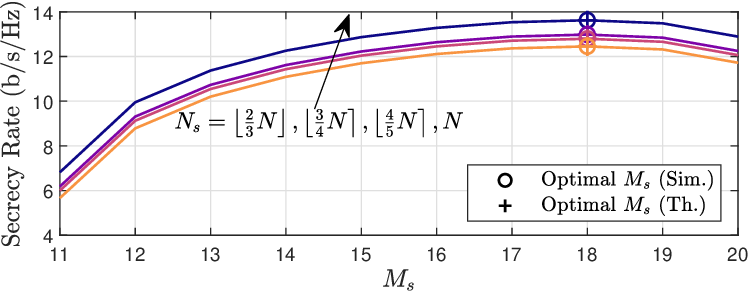}
		\caption{\emph{Worst-case secrecy rate}}
		\label{fig:optms1} 
	\end{subfigure}
	\vfill
	\begin{subfigure}[b]{0.5\textwidth}
		\centering
		\includegraphics[width=0.8\linewidth]{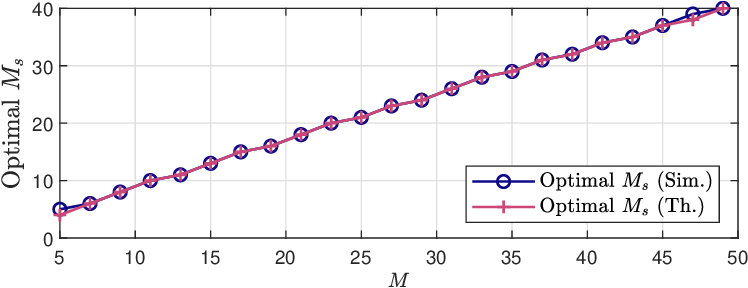}
		\caption{Optimal $M_s$ with various $M$, $N_s = N$}
		\label{fig:optms2} 
	\end{subfigure} 
	\caption{{Performance of the optimized {RIBES}, $\theta_{\mathrm{B}} = \theta_{\mathrm{E}}$}}
	\label{fig:optms} \vspace{-1em}
\end{figure}

\figref{fig:optms1} illustrates the \emph{worst-case secrecy rate} over the parameter $M_s$ when Eve is positioned within the wiretap area. Moreover, we compare the optimal value of $M_s$ that maximizes the \emph{worst-case secrecy rate} obtained through simulation and the theoretical optimal value obtained by \eqref{eq:opt_M_s} in {Theorem} \ref{theorem:opt_M_s_N_s}. The perfect match between the two indicates that \eqref{eq:opt_M_s} in {Theorem} \ref{theorem:opt_M_s_N_s} is accurate. Furthermore, it is noteworthy that the optimal value of $M_s$ does not depend on the value of $N_s$ {in these results}. This is due to the fact that the received SNR at Bob is significantly larger $\SI{0}{dB}$, rendering the optimal value of $M_s$ independent of $N_s$ (as per {Proposition} \ref{proposition:opt_M_s_N_s}). In addition, we present the optimal $M_s$ as a function of $M$ in \figref{fig:optms2}. The figure shows that the optimal $M_s$ computed by \eqref{eq:opt_M_s} in {Theorem} \ref{theorem:opt_M_s_N_s}, closely matches the results obtained through simulation with varying $M$.

\begin{figure}
	\centering
	\includegraphics[width=0.825\linewidth]{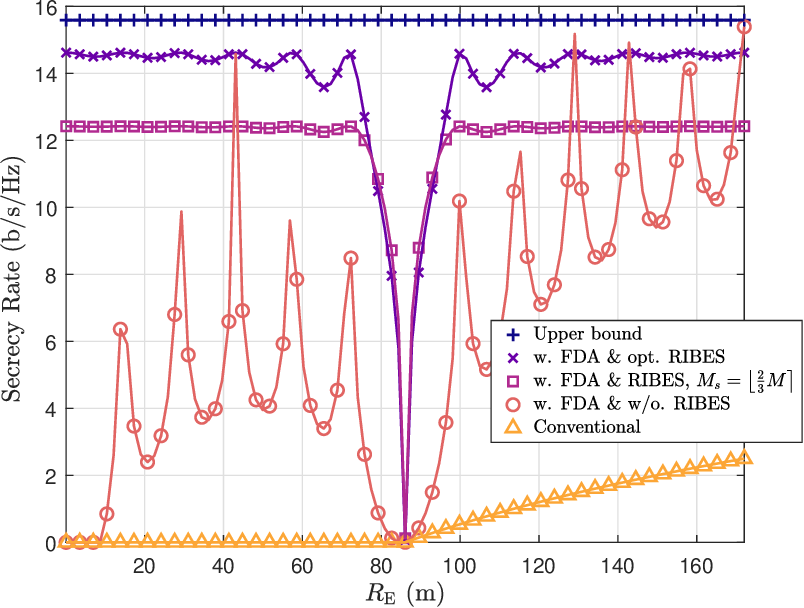}
	\caption{Secrecy rate with respect to $R_{\mathrm{E}}$. Markers: simulation results;
		curves: theoretical results.}
	\label{fig:secrecyratetraj1m21test} \vspace{-1.5em}
\end{figure}
In \figref{fig:secrecyratetraj1m21test}, we illustrate the secrecy rate as a function of $R_{\mathrm{E}}$. Thereby, the results of different transmission techniques are compared, including the employment of a conventional phased array at the BS, using FDA at the BS without {RIBES}, leveraging FDA and {RIBES} jointly with $M_s = \left\lfloor \frac{2}{3} M \right\rceil$, and using FDA along with optimized {RIBES}. From this figure, we can observe that the simulation results align with the analytical results, verifying the accuracy of the closed-form expressions derived in Sec. \ref{sec:secrecy}. { Furthermore, when $R_{\mathrm{E}} < R_{\mathrm{B}}$, the secrecy rate is zero when a conventional phased array is employed. }However, when employing FDA without {RIBES}, a positive secrecy rate can be achieved, in cases where $R_{\mathrm{E}} > \SI{10}{\meter}$. Further, when FDA and {RIBES} are applied jointly, we observe a high secrecy rate even if Eve is at a very short distance from the RIS. This demonstrates the robustness of {RIBES} against high channel gain at Eve. Additionally, when Eve is in proximity to Bob, the secrecy rate achieved with both FDA and {RIBES} surpasses the secrecy rate obtained when only FDA is applied. Moreover, we see that the secrecy performance can be enhanced when optimized {RIBES} is applied. It is noted that we also observe a higher secrecy rate without {RIBES} compared to with {RIBES} at a few specific values of $R_{\mathrm{E}}$. This is because, at these specific values, the eavesdropper is either close to or positioned at the null points. In such cases, the upper bound of the secrecy rate specified in \eqref{eq:sec_rate_UB} is achieved at the null points without {RIBES}.
\subsection{{Equal distance from the RIS at Bob and Eve} }
Now, we evaluate the performance of the proposed transmission technique when  $R_{\mathrm{B}}= R_{\mathrm{E}}$. Thereby, we set $\Delta F = 0$, since FDA does not confer any advantages. This is because we have $\mathbb{V}[u] = 0$ according to  \lemmaref{lemma:lemma2} \eqref{eq:variance_u}. 

\begin{figure}
	\centering
	\includegraphics[width=0.8\linewidth]{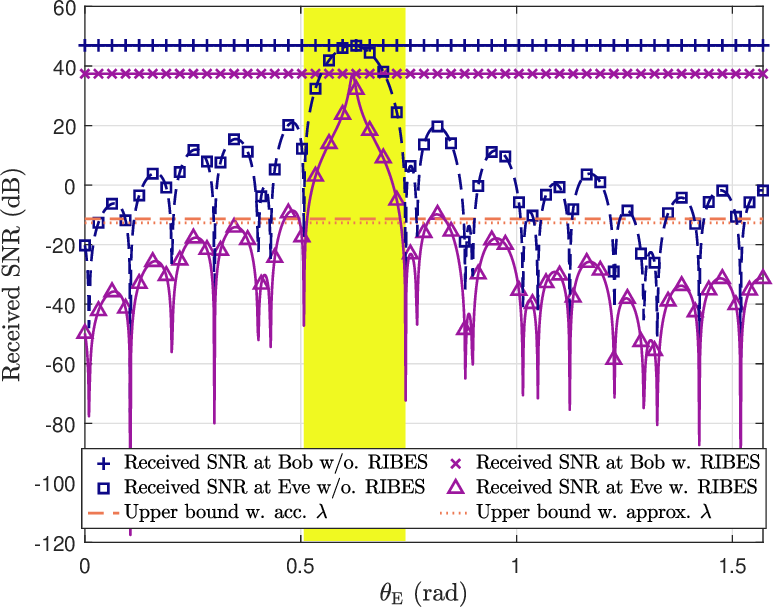}
	\caption{Comparison of the received SNR at Bob and Eve with different transmission techniques,  $N_s = \left \lfloor \frac{2}{3} N \right \rceil $ and $M_s = M$. Markers: simulation results;
		curves: theoretical results.}
	\label{fig:receivedsnrvsubtraj2} \vspace{-1.5em}
\end{figure}

In \figref{fig:receivedsnrvsubtraj2}, we compare the received SNR at Bob and Eve with and without {RIBES}. We observe that the markers align well with the curves, confirming the correctness of the closed-form expressions. 
 Meanwhile, when Eve is located in the target area (indicated with a yellow background), and as the distance between the AOA at Bob and Eve increases, the received SNR at Eve decreases more rapidly with {RIBES} than without it. In addition, we compare the upper bound of the received SNR at Eve in the wiretap area when using the accurate $\lambda$ and the approximate $\lambda$ observed by \eqref{eq:appro_lambda} in \lemmaref{lemma:lemma2}. Here, the accurate $\lambda$ is observed by simulations.  The {corresponding} curves show that there is only a slight difference between the upper bounds computed by accurate and approximate $\lambda$. Therefore, it is reasonable to utilize the approximate upper bound for determining the optimal value of $N_s$. 
\begin{figure}
	\centering
	\begin{subfigure}[b]{0.5\textwidth}
	\centering
	\includegraphics[width=0.8\linewidth]{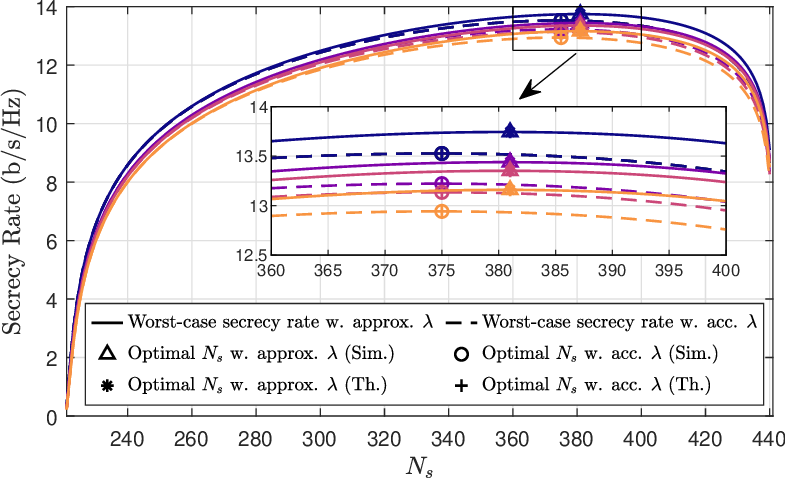}
	\caption{\emph{Worst-case secrecy rate}. Blue: $M_s =  M$; purple: $M_s = \left \lfloor \frac{4}{5} M \right \rceil $; red: $M_s = \left \lfloor \frac{3}{4} M \right \rceil $; orange: $M_s = \left \lfloor \frac{2}{3} M \right \rceil $}
	\label{fig:secrecyratens} 
\end{subfigure}
\vfill
\begin{subfigure}[b]{0.5\textwidth}
	\centering
	\includegraphics[width=0.8\linewidth]{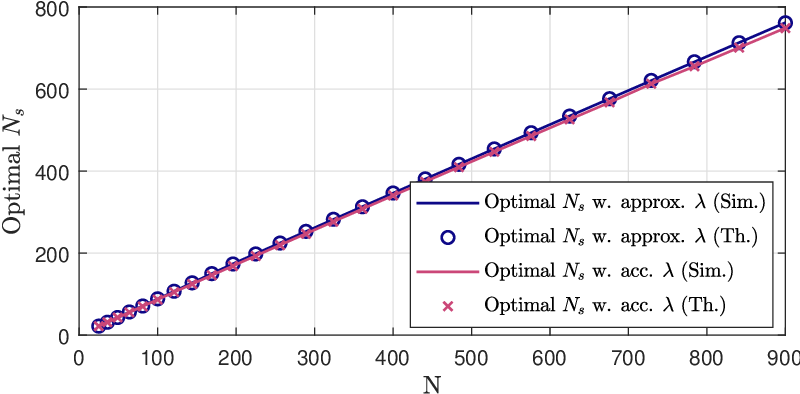}
	\caption{Optimal $N_s$ with various $N$, $M_s = M$}
	\label{fig:optNs}  
\end{subfigure} 
	\caption{{Performance of the optimized {RIBES}, $R_{\mathrm{B}} = R_{\mathrm{E}}$}} \vspace{-1em}
\end{figure}

Now, we illustrate the \emph{worst-case secrecy rate} in the wiretap area as a function of $N_s$ in \figref{fig:secrecyratens}. In particular, we compare the upper bound with approximate and accurate $\lambda$, where we observe a minor gap between them. Nevertheless, these two bounds exhibit a similar behavior across various values of $N_s$. In addition, the value of $N_s$ optimized using the approximate $\lambda$ is close to that using accurate $\lambda$. This can be observed from \figref{fig:optNs}, where we plot the optimal $N_s$ as function of $N$. In both figures, we observe a perfect match between the optimal $N_s$ obtained through simulation and the closed-form expression in \eqref{eq:optimal_N_s_appr}, which verifies the validity of {Theorem} \ref{theorem:opt_M_s_N_s}. Meanwhile, from \figref{fig:secrecyratens}, the optimal $N_s$ remains unchanged across different values of $M_s$ due to the fact that $\gamma_\mathrm{B}^\star \gg 1$ (as seen in {Proposition} \ref{proposition:opt_M_s_N_s}).

\begin{figure}
	\centering
	\includegraphics[width=0.8\linewidth]{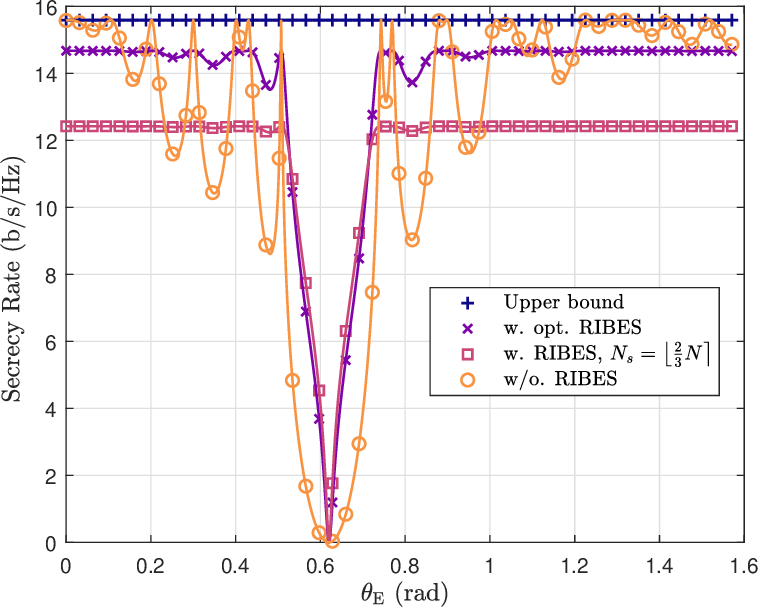}
	\caption{Secrecy rate as a function of $\theta_{\mathrm{E}}$, $M_s = M$. Markers: simulation results;
		curves: theoretical results.}
	\label{fig:secrecyratetraj2m21n21} \vspace{-1.5em}
\end{figure}
In \figref{fig:secrecyratetraj2m21n21}, we plot the secrecy rate as a function of $\theta_{\mathrm{E}}$. It can be observed that when Eve is located close to Bob, i.e., when the distance in the AOA between Bob and Eve is small, the secrecy performance with {RIBES} significantly outperforms that without {RIBES}. As this distance increases, the secrecy rate with {RIBES} converges to a constant, while the secrecy rate without {RIBES} remains dependent on the AOA. Meanwhile, at some specific values of $\theta_{\mathrm{E}}$, the secrecy rate without {RIBES} is higher than with {RIBES}. This occurs because the received SNR at Eve is close to or equals zero at these particular values, and the upper bound of the secrecy rate is achieved when {RIBES} is not employed according to \eqref{eq:sec_rate_UB}. Furthermore, it is evident that optimizing the {RIBES} can enhance the secrecy rate and generally results in a better overall performance compared to not using {RIBES}. \vspace{-1em}
\subsection{Eve on a circle around Bob}
We now examine the secrecy rate when Eve is located with a fixed distance to Bob. In this case, the locations of Eve can be determined by $x_{\mathrm{E}} = x_{\mathrm{B}}+ D \cos \varphi$ and $y_{\mathrm{E}} = y_{\mathrm{B}} + D \sin \varphi$, for $\varphi \in [0, 2\pi]$, where $D$ is the radius of the trajectory.  In \figref{fig:secrecyratetraj3m21}, we depict the secrecy rate as a function of $\varphi$. It is evident that the markers closely match the curves, which once again verifies the accuracy of the closed-form expressions. In addition, we can see that when $\varphi$ ranges between 2 and 3 radians, a zero secrecy rate is obtained when conventional phases array is applied. This is because in this range, Eve is almost aligned with Bob ($\theta_{\mathrm{B}} \approx \theta_{\mathrm{E}} $) and Eve is closer to the RIS than Bob. When FDA is utilized, a positive secrecy rate can be always achieved. However, we still observe a low secrecy rate when Eve is located in the same direction as Bob without {RIBES}. From this figure, we can see that the utilization of {RIBES} enhances the secrecy rate, especially when Eve is aligned with Bob. Furthermore, the proposed optimized {RIBES} can effectively improve the secrecy rate. 
\begin{figure}
	\centering
	\includegraphics[width=0.8\linewidth]{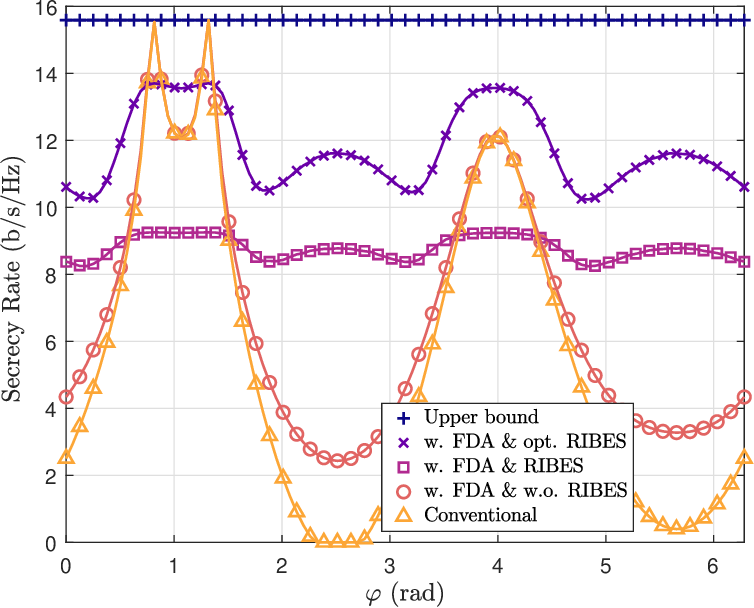}
	\caption{Secrecy rate with $D = \SI{10}{\meter}$. Markers: simulation results;
		curves: theoretical results.}
	\label{fig:secrecyratetraj3m21} \vspace{-1.5em}
\end{figure}

\section{Conclusion}
\label{sec:conclusion}
In this work, we have proposed FDA with {RIBES} to enable {secure communication in an RIS-assisted mmWave/THz communication system, providing both direction and range protection.} Specifically, we have considered a passive eavesdropper and designed the transmit beamforming and RIS configuration based on Bob's channel information. We have further derived closed-form expressions for the secrecy rate using the proposed algorithms, and these results were validated through extensive simulations. Additionally, the \emph{worst-case secrecy rate} has been analyzed for the case that Eve is located in the wiretap area. Furthermore, we have determined the optimal subset sizes of the selected transmit antennas and reflective elements in closed form. Numerical results demonstrate that the proposed algorithm provides {an enhanced secrecy rate} compared to conventional approaches, particularly when Eve is in the proximity to Bob or the RIS. 
 \section*{Appendix}
  \subsection{Proof of Lemma \ref{lemma:lemma1}}	
  {By substituting \eqref{eq:opt_IRS_closed_form_2} in to \eqref{eq:h_UE_m}, the cascaded channel of Bob is computed as
  \begin{align}
  		%	\label{eq:h_bob_c}
  		&\mathbf{h}_{\mathrm{B},m }^\star(k) 
  		%	= \sqrt{L_\mathrm{G}(R_1) L_H(R_{\mathrm{B}})} e^{-j2\pi f_m \left(  \frac{R_1+R_{\mathrm{B}}}{c} + \tilde{\tau}_{m} \right) } \nonumber \\ & \qquad \times {\sum_{n = 1}^{N} e^{-j 2\pi f_m \tilde{\tau}_{n}^{\mathrm{B}}} e^{j \phi_n^\star(k)} } \nonumber \\ &
  		=\sqrt{L_\mathrm{G}(R_1) L_H(R_{\mathrm{B}})} e^{-j2\pi f_m \left(  \frac{R_1+R_{\mathrm{B}}}{c} + \tilde{\tau}_{m} \right) } \times \nonumber \\ & {\left( \textstyle \sum_{n \in \mathcal{N}_s(k)} e^{-j 2\pi f_m \tilde{\tau}_{n}^{\mathrm{B}}} e^{j \phi_n^\star(k)} \!+\! \textstyle \sum_{n \notin \mathcal{N}_s(k)} e^{-j 2\pi f_m \tilde{\tau}_{n}^{\mathrm{B}}} e^{j \phi_n^\star(k)} \! \right) } \nonumber \\ &
  		\overset{(a)}{=}\sqrt{L_\mathrm{G}(R_1) L_\mathrm{H}(R_{\mathrm{B}})} e^{-j2\pi f_m \left(  \frac{R_1+R_{\mathrm{B}}}{c} + \tilde{\tau}_{m} \right) } \left( N_s -\left( N \! - \! N_s\right) \right) \nonumber \\ &
  		= \sqrt{L_\mathrm{G}(R_1) L_\mathrm{H}(R_{\mathrm{B}})} e^{-j2\pi f_m \left(  \frac{R_1+R_{\mathrm{B}}}{c} + \tilde{\tau}_{m} \right) } (2 N_s -N), \nonumber
  \end{align}
where (a) holds as ${2 \pi \Delta f_m \tilde{\tau}^{\mathrm{B}}_n}$ is significantly smaller compared to ${2 \pi \Delta f_0 \tilde{\tau}^{\mathrm{B}}_n}$, and thus, it is disregarded. }
Accordingly, we have 
  $\left\| {\mathbf{h}}_{\mathrm{B} }^\star(k)\right\| = \sqrt{M {L_\mathrm{G}(R_1) L_\mathrm{H}(R_{\mathrm{B}})}} (2 N_s -N). $
  As a result, the signal received by Bob can be expressed as
{\begin{align}
%  		\label{eq:y_bob_star}
  		&y_{\mathrm{B}}^\star(k) = 	\sqrt{P} {\mathbf{h}}_{\mathrm{B} }^\star(k)^H \boldsymbol{w}^{\star}(k) x(k) + n_{\mathrm{B}}(k) \nonumber \\ 
  		&= \frac{\sqrt{P}x(k)}{ \left\| {\mathbf{h}}_{\mathrm{B} }^\star(k)\right\| }{\mathbf{h}}_{\mathrm{B} }^\star(k)^H \left( {\mathbf{h}}_{\mathrm{B} }^\star(k) \circ \mathbf{a}(k) \right) +   n_{\mathrm{B}}(k)   \nonumber \\ 
  		&= \!\! \frac{\sqrt{P}\! x(k) \! }{ \left\| {\mathbf{h}}_{\mathrm{B} }^\star\! (k) \! \right\| } \! \left( \!  \textstyle  \sum_{m \in \mathcal{M}_s(k)} \! \left| {\mathbf{h}}_{\mathrm{B},m }^\star\right|^2 \!\!-\! \!\!  \textstyle \sum_{m \notin \mathcal{M}_s(k)} \!  \left| {\mathbf{h}}_{\mathrm{B},m}^\star \right|^2 \! \right) \! \!+\! n_{\mathrm{B}}(k) \nonumber \\ 
  		& = \frac{\sqrt{P}x(k) L_\mathrm{G}(R_1) L_\mathrm{H}(R_{\mathrm{B}}) (2 N_s -N)^2}{ \sqrt{M {L_\mathrm{G}(R_1) L_\mathrm{H}(R_{\mathrm{B}})}} (2 N_s -N) } (2M_s\!-\!M)\!+\!   n_{\mathrm{B}}(k) \nonumber \\ 
  		&= \sqrt{\frac{P L_\mathrm{G}(R_1) L_\mathrm{H}(R_{\mathrm{B}})}{M}}  (2 M_s\!-\!M)(2N_s\!-\!N)x(k) +  n_{\mathrm{B}}(k). \nonumber
  \end{align}}
  %Here, we need ensure $2 M_s-M$
%  \textcolor{orange}{ Note that, in this work, we ensure that Bob can always benefit from transmit-beamforming and RIS by specifying $M_s > M - M_s$ and $N_s > N - N_s$, i.e., $M_s > \frac{M}{2}$ and $N_s>\frac{N}{2}$.}  
  Thereby, the received SNR at Bob with {RIBES} can be computed as
  \begin{align}
  	\label{eq:SNR_Bob_FDA_w_sel}
  	\gamma_{\mathrm{B}}^\star =  \frac{P}{\sigma_{\mathrm{B}}^2} L_\mathrm{G}(R_1) L_\mathrm{H}(R_{\mathrm{B}}) \frac{(2 M_s-M)^2}{M}(2N_s-N)^2.
  \end{align}
 \subsection{Proof of Lemma \ref{lemma:lemma2}}	
 \label{sec:proof1}
As $u$ and $v$ are two independent and uncorrelated parameters, the expectation of the scaling factor is computed by
\begin{align}
	\label{eq:proof_mean_beta}
	\mathbb{E}[\beta  (R_{\mathrm{E}}, \theta_{\mathrm{E}})] =  
	\sqrt{ \frac{ L_\mathrm{G}(R_1)L_\mathrm{H}(R_\mathrm{E})}{M}} 	\mathbb{E} [u] 	\mathbb{E}[v].
\end{align}
Further, the mean value of $u$ is calculated as
\begin{align}
	\label{eq:proof_mean_u}
	& \mathbb{E} [u] \nonumber \\
	& =  	\mathbb{E}\Big[ 2 \textstyle \sum_{m \in \mathcal{M}_s(k)} \!e^{j 2\pi 
		\Delta	f_m \frac{R_{\mathrm{E}}-R_{\mathrm{B}}}{c}} \!\! - \!\! \textstyle \sum_{m =1}^{M}\! e^{j 2\pi 
		\Delta	f_m  \frac{R_{\mathrm{E}}-R_{\mathrm{B}}}{c}}\Big] \nonumber \\
	& = 2 	\mathbb{E}\Big[ \textstyle \sum_{m \in \mathcal{M}_s(k)} \!e^{j 2\pi 
		\Delta	f_m \frac{R_{\mathrm{E}}-R_{\mathrm{B}}}{c}}\Big]\! - \!\textstyle \sum_{m =1}^{M}\! e^{j 2\pi 
		\Delta	f_m  \frac{R_{\mathrm{E}}-R_{\mathrm{B}}}{c}} \nonumber \\
	& \stackrel{(a)}{=} 2 \frac{M_s}{M} \textstyle \sum_{m =1}^{M}\! e^{j 2\pi 
		\Delta	f_m  \frac{R_{\mathrm{E}}-R_{\mathrm{B}}}{c}} -\textstyle \sum_{m =1}^{M}\! e^{j 2\pi 
		\Delta	f_m  \frac{R_{\mathrm{E}}-R_{\mathrm{B}}}{c}} 
	\nonumber \\
	& = \frac{2M_s-M}{M}  \textstyle \sum_{m =1}^{M}\! e^{j 2\pi 
		\Delta	f_m  \frac{R_{\mathrm{E}}-R_{\mathrm{B}}}{c}}    \nonumber \\
	& = \frac{2M_s-M}{M}\frac{e^{-j \pi 
			M\Delta F \frac{R_{\mathrm{E}}-R_{\mathrm{B}}}{c}}-e^{j \pi 
			M\Delta F \frac{R_{\mathrm{E}}-R_{\mathrm{B}}}{c}}}{e^{-j \pi 
			\Delta F \frac{R_{\mathrm{E}}-R_{\mathrm{B}}}{c}}-e^{j \pi 
			\Delta F \frac{R_{\mathrm{E}}-R_{\mathrm{B}}}{c}}}	
	\nonumber \\
	& = \frac{2M_s-M}{M} 	\underbrace{\frac{ \sin \left(M \pi\Delta F (R_{\mathrm{E}}-R_{\mathrm{B}}) /c\right)  }{\sin \left( \pi\Delta F (R_{\mathrm{E}}-R_{\mathrm{B}}) /c \right) }}_{\mu_1},
\end{align}
where (a) follows the expectation of random sampling without replacement \cite{horvitz1952generalization}. 
%where 
%\begin{align}
%\mu_1 &= \sum_{m =0}^{M-1}\! e^{j 2\pi 
	%	\Delta	f_m  \frac{R_{\text{Eve}}-R_{\text{Bob}}}{c}} 	\nonumber \\
%& = e^{j 2\pi 
%	\Delta F	(-\frac{M-1}{2})  \frac{R_{\text{Eve}}-R_{\text{Bob}}}{c}} \frac{1-e^{j 2\pi 
	%	M\Delta F \frac{R_{\text{Eve}}-R_{\text{Bob}}}{c}} }{1-e^{j 2\pi 
	%	\Delta F \frac{R_{\text{Eve}}-R_{\text{Bob}}}{c}}}\nonumber \\
%& = \frac{e^{-j \pi 
%		M\Delta F \frac{R_{\text{Eve}}-R_{\text{Bob}}}{c}}-e^{j \pi 
%		M\Delta F \frac{R_{\text{Eve}}-R_{\text{Bob}}}{c}}}{e^{-j \pi 
%		\Delta F \frac{R_{\text{Eve}}-R_{\text{Bob}}}{c}}-e^{j \pi 
%		\Delta F \frac{R_{\text{Eve}}-R_{\text{Bob}}}{c}}}
%\nonumber \\
%& =	\frac{ \sin \left(M \pi\Delta F (R_{\text{Eve}}-R_{\text{Bob}}) /c\right)  }{\sin \left( \pi\Delta F (R_{\text{Eve}}-R_{\text{Bob}}) /c \right) }.
%\end{align}
Similarly, the mean value of $v$ can be computed by
\begin{align}
\label{eq:proof_mean_v}
& \mathbb{E} [v] = 2 \mathbb{E} \Big[   \textstyle \sum_{n \in \mathcal{N}_s(k)} e^{j 2\pi 
f_0 \left( \tilde{\tau}_n^{\mathrm{E}} \! - \! \tilde{\tau}_n^{\mathrm{B}} \right) }\Big] \! - \! \textstyle \sum_{n =1}^{N} e^{j 2\pi 
f_0 \left( \tilde{\tau}_n^{\mathrm{E}} - \tilde{\tau}_n^{\mathrm{B}} \right) } \nonumber \\
& = \frac{2N_s-N}{N} \textstyle \sum_{n =1}^{N} e^{j 2\pi 
f_0 \left( \tilde{\tau}_n^{\mathrm{E}} - \tilde{\tau}_n^{\mathrm{B}} \right) }\nonumber \\
& = \frac{2N_s-N}{N}  \textstyle \sum_{n_\mathrm{H} =1}^{N_\mathrm{H}}  \sum_{n_\mathrm{V}=1}^{N_\mathrm{V}} e^{j 2\pi 
f_0 \left( \tilde{\tau}_n^{\mathrm{E}} - \tilde{\tau}_n^{\mathrm{B}} \right) }\nonumber \\
& = \frac{2N_s-N}{N}    \textstyle \sum_{n_\mathrm{H} =1}^{N_\mathrm{H}} e^{j 2\pi 
\frac{f_0}{c} d_\mathrm{H} ( \cos(\theta_{\mathrm{B}})-\cos(\theta_{\mathrm{E}})) (n_\mathrm{H}-\frac{N_\mathrm{H}+1}{2}) } \nonumber \\
& \quad \times \textstyle \sum_{n_\mathrm{V} =1}^{N_\mathrm{V}}   e^{j 2\pi 
\frac{f_0}{c} d_\mathrm{V} ( \sin(\theta_{\mathrm{E}})-\sin(\theta_{\mathrm{B}})) (n_\mathrm{V}-\frac{N_\mathrm{V}+1}{2}) } 
\nonumber \\
& = \frac{2N_s-N}{N} \underbrace{\frac{ \sin \left( 0.5 N_\mathrm{H} \pi  ( \cos(\theta_{\mathrm{B}})-\cos(\theta_{\mathrm{E}}))\right)  }{\sin \left( 0.5 \pi   ( \cos(\theta_{\mathrm{B}})-\cos(\theta_{\mathrm{E}})) \right) }}_{\mu_2} 	\nonumber \\
& \quad \times \underbrace{\frac{ \sin \left( 0.5 N_\mathrm{V} \pi  ( \sin(\theta_{\mathrm{E}})-\sin(\theta_{\mathrm{B}}))\right)  }{\sin \left( 0.5 \pi  ( \sin(\theta_{\mathrm{E}})-\sin(\theta_{\mathrm{B}})) \right) }}_{\mu_3}.
\end{align}
By substituting \eqref{eq:proof_mean_u} and \eqref{eq:proof_mean_v}  into \eqref{eq:proof_mean_beta}, we obtain \eqref{eq:mean_scaling_fator} in Lemma 1.
%	\begin{align}
%	\label{eq:mean_scaling_fator2}
%	&\mathbb{E}[\beta  (R_{\text{Eve}}, \theta_{\text{Eve}})]  \nonumber \\= &\frac{1}{\sqrt{M}}  \frac{C_0}{\sqrt{ R_1^{\alpha_1}R_{\text{Eve}}^{\alpha_2}}}  
%	\frac{\left( 2M_s-M\right) \left( 2N_s-N \right) }{MN}  \mu_1 \mu_2 \mu_3.
%\end{align}	
Furthermore, for the variance of the scaling factor, we have 
\begin{align}
	&\mathbb{V}[\beta  (R_{\mathrm{E}}, \theta_{\mathrm{E}})] = \frac{1}{{M}}  L_\mathrm{G} (R_1) L_\mathrm{H}(R_\mathrm{E})\mathbb{V} [u v]  \nonumber \\
	=&\frac{1}{{M}}  L_\mathrm{G} (R_1) L_\mathrm{H}(R_\mathrm{E}) \left(	\mathbb{V}[u] \mathbb{V}[v] \!+\!	\mathbb{V}[v] \left| \mathbb{E}[u]\right| ^2 \! + \! \mathbb{V}[u] \left| \mathbb{E}[v]\right| ^2 \right).\nonumber
\end{align}
Given the fact that $\mathbb{V} [x] = \mathbb{E} [x^2] -  \mathbb{E} [x]^2$, and let $x = \sum_{m \in \mathcal{M}_s(k)} \!e^{j 2\pi 
\Delta	f_m \frac{R_{\mathrm{E}}-R_{\mathrm{B}}}{c}} $, the variance of $u$ is calculated as 
\begin{align}
\mathbb{V}[u]&= 4 	\mathbb{V} \left[ \textstyle \sum_{m \in \mathcal{M}_s(k)} \!e^{j 2\pi 
\Delta	f_m \frac{R_{\mathrm{E}}-R_{\mathrm{B}}}{c}}\right] 	= 4  \left( \mathbb{E} [x^2] -  \mathbb{E} [x]^2\right),\nonumber %	\nonumber \\
%	&= 4 \left( M_s + \frac{M_s-M}{M-1} \left( \left( \sum_{m =0}^M \!e^{j 2\pi 
%		\Delta	f_m \frac{R_{\text{Eve}}-R_{\text{Bob}}}{c}}\right)  ^2 -M\right) \right) 
\end{align}	
where we have 
\begin{align}
\mathbb{E} [x^2] &=  M_s + \frac{M_s-1}{M-1} \left( \left( \textstyle \sum_{m =1}^M \!e^{j 2\pi 
\Delta	f_m \frac{R_{\mathrm{E}}-R_{\mathrm{B}}}{c}}\right)  ^2 -M\right) \nonumber \\ 
&= M_s + \frac{M_s-1}{M-1} \big( \mu_1  ^2 -M\big), \nonumber
\end{align}	
and 
$
\mathbb{E} [x]^2 =   \frac{M_s^2}{M^2} \mu_1^2.
$
Thereby, we obtain \begin{align}
\label{eq:proof_var_u}
\mathbb{V}[u] &= 4 \left(  M_s + \frac{M_s-1}{M-1} \left( \mu_1  ^2 -M\right) -\frac{M_s^2}{M^2} \mu_1^2  \right) \nonumber \\ &=  4 \frac{M_s(M-Ms)}{M(M-1)}\left(M- \frac{1}{M} \mu_1^2\right).
\end{align}	
The variance of $v$ is obtained similarly.

%\begin{align}
%\mathbb{V}[v] &= 4 \frac{N_s(N-Ns)}{N(N-1)}(N- \frac{1}{N} \mu_2^2 \mu_3^2)\mathbb{V}[v].
%\end{align}	
 \subsection{Proof of Lemma \ref{lemma:upper_bound}}
  \label{sec:proof2}
We first prove the upper bound of the received SNR at Eve when $\theta_{\mathrm{E}} = \theta_{\mathrm{B}}$. In this case, we have $\mu_2 = N_\mathrm{H}$, $\mu_3 = N_\mathrm{V}$, and $\mathbb{V}[v]  = 0$.  From Lemma 1, we obtain the mean 
 \begin{align}
 	\label{eq:mean_beta_equal_theta}
 	&\mathbb{E}[\beta  (R_{\mathrm{E}}, \theta_{\mathrm{E}})] \nonumber \\
 	 &= \frac{\mu_1 }{\sqrt{M}}  \sqrt{L_\mathrm{G}(R_1)L_\mathrm{H}(R_\mathrm{E})}  \frac{\left( 2M_s\!-\!M\right) \left( 2N_s\!-\!N \right) }{M},
 \end{align}
while the variance is 
 \begin{align}
\label{eq:var_beta_equal_theta}
& \mathbb{V}[\beta  (R_{\mathrm{E}}, \theta_{\mathrm{E}})] = \frac{1}{{M}}  L_\mathrm{G}(R_1) L_\mathrm{H}(R_\mathrm{E})\mathbb{V}[u] \left| \mathbb{E}[v]\right| ^2 \\
 & =   L_\mathrm{G}(R_1) L_\mathrm{H}(R_\mathrm{E})4 \frac{M_s(M \! - \! Ms)}{M^2(M \! - \!1)}(M \!-\! \frac{1}{M} \mu_1^2) (2N_s \!-\! N)^2.  \nonumber
  \end{align}
By substituting \eqref{eq:mean_beta_equal_theta} and \eqref{eq:var_beta_equal_theta} into \eqref{eq:SNR_Eve_w_random_Sel}, we observe
 \begin{align}
 	\label{eq:gamma_E_UB_proof}
 	&\gamma_{E}^\star \nonumber  \\ 
 	&=\! \frac{P \frac{\mu_1^2 }{M} L_\mathrm{G}(R_1) L_\mathrm{H}(R_\mathrm{E}) 
 		\frac{\left( 2M_s-M\right)^2 \left( 2N_s-N \right)^2 }{M^2} }{PL_\mathrm{G}(R_1) L_\mathrm{H}(R_\mathrm{E}) 4 \frac{M_s(M\!-\!Ms)}{M^2(M\!-\!1)}(M \! -\! \frac{1}{M} \mu_1^2) (2N_s\!-\!N)^2\!+\!\sigma_{\mathrm{E}}^2} \nonumber  \\
 	& \leqslant   \frac{P \frac{\mu_1^2 }{M}   L_\mathrm{G}(R_1) L_\mathrm{H}(R_\mathrm{E})
 		\frac{\left( 2M_s-M\right)^2 \left( 2N_s-N \right)^2 }{M^2} }{P  L_\mathrm{G}(R_1) L_\mathrm{H}(R_\mathrm{E}) 4 \frac{M_s(M-Ms)}{M^2(M-1)}(M- \frac{1}{M} \mu_1^2) (2N_s-N)^2}\nonumber 
 	\\
 	 & = \frac{\mu_1^2 \left( 2M_s-M\right)^2 (M-1)}{4 M_s (M-M_s) M(M-\frac{1}{M} \mu_1^2)} \nonumber  \\
 	% & \approx \frac{\mu_1^2 \left( 2M_s-M\right)^2 (M-1)}{4 M_s (M-M_s) M^2}  \nonumber  \\
 	 &  < \frac{ \max_{{R_{\mathrm{E}}} }(\mu_1^2 ) \left( 2M_s-M\right)^2 (M-1)}{4 M_s (M-M_s) (M^2-\max_{R_{\mathrm{E}}} (\mu_1^2 ) )}.
 	  \end{align}
   Here, if Eve is within the wiretap area, the maximum value of $\mu_1^2$ occurs when $2\pi M \Delta F \frac{R_{\mathrm{E}} - R_{\mathrm{B}}}{c} = \pm 3\pi$. This point is exactly the midpoint between the first and second null of $\mu_1$. 
   Thus, we have
    \begin{align}
    	\label{eq:max_mu_1}
    \max_{{(\theta_{\mathrm{E}}, R_{\mathrm{E}}) \in	\mathcal{R}} } \mu_1^2 = \left( \frac{\sin( \pm \frac{3 \pi}{2})}{\sin(\pm \frac{3}{2 M } \pi)}\right) ^2 = \frac{1}{\sin^2( \frac{3 \pi}{2 M })}.	
    \end{align}
Next, by substituting  \eqref{eq:max_mu_1} in \eqref{eq:gamma_E_UB_proof}, we observe \eqref{eq:UB1} in Lemma \ref{lemma:upper_bound}.  Similarly,  if $R_{\mathrm{E}} = R_{\mathrm{B}}$, we have 
 \begin{align}
	\label{eq:gamma_E_UB2_proof}
	\gamma_{E}^\star 
	  &< \frac{  \left( 2N_s-N \right)^2 (M-1) \max_{\theta_{\mathrm{E}}}(\mu_2^2 \mu_3^2)}{4 N_s (N-N_s) (N^2-\max_{\theta_{\mathrm{E}}}(\mu_2^2 \mu_3^2))} \nonumber \\
	  & = \frac{ \left( 2N_s-N \right)^2 (M-1) \lambda^2 }{4 N_s (N-N_s) (N^2-\lambda^2)} ,
\end{align}
where $\lambda = \max_{\theta_{\mathrm{E}}} \left| \mu_2 (\theta_{\mathrm{E}}) \mu_3 (\theta_{\mathrm{E}})\right| $, for ${(\theta_{\mathrm{E}}, R_{\mathrm{E}}) \in	\mathcal{R}} $.
However, an accurate closed-form for $\lambda$ is challenging to observe. Thus, we approximate $\lambda$ as  
$
 	\lambda = \max   (   \lambda_1, \lambda_2)
$,
{where $\lambda_1$  and $\lambda_2$ are obtained as $|\mu_2 (\hat{\theta}_{\mathrm{E}})| |\mu_3(\hat{\theta}_{\mathrm{E}})|$ and $|\mu_2 (\tilde{\theta}_{\mathrm{E}})| |\mu_3(\tilde{\theta}_{\mathrm{E}})|$, respectively. Here, $|\mu_2(\theta_{\mathrm{E}})|$ and $|\mu_3(\theta_{\mathrm{E}})|$ achieve their maxima at $\tilde{\theta}_{\mathrm{E}}$ and $\hat{\theta}_{\mathrm{E}}$, respectively.} 
Similar to \eqref{eq:max_mu_1}, we obtain $ \max(|\mu_2|) = \frac{1}{\sin (\frac{3 \pi}{2 N_\mathrm{H}})}$, which is observed when $N_\mathrm{H} \pi  (\cos\theta_{\mathrm{B}}-\cos\theta_{\mathrm{E}}) = \pm 3\pi$. As a result, we obtain $\theta_{\mathrm{E}} = \arccos (\cos\theta_{\mathrm{B}} \pm \frac{3}{N_\mathrm{H}})$. Thereby, $\lambda_1 = \frac{1}{\sin (\frac{3 \pi}{2 N_\mathrm{H}})} \frac{ \sin \left( 0.5 N_\mathrm{V} \pi  ( \sin\theta_{\mathrm{E}}-\sin\theta_{\mathrm{B}})\right)  }{\sin \left( 0.5 \pi  ( \sin\theta_{\mathrm{E}}-\sin\theta_{\mathrm{B}}) \right) } = \frac{ \sin \left( 0.5 N_\mathrm{V} \pi  ( \sqrt{1-(\cos{\theta_{\mathrm{B}}} \pm \frac{3}{N_\mathrm{H}})^2}-\sin\theta_{\mathrm{B}})\right)  }{\sin(\frac{3 \pi}{2 N_\mathrm{H}})\sin \left( 0.5 \pi  ( \sqrt{1-(\cos{\theta_{\mathrm{B}}} \pm \frac{3}{N_\mathrm{H}})^2}-\sin\theta_{\mathrm{B}})\right) } $, as we have $\sin(\arccos (\cos\theta_{\mathrm{B}} \pm \frac{3}{N_\mathrm{H}}) ) = \sqrt{1-(\cos\theta_{\mathrm{B}} \pm \frac{3}{N_\mathrm{H}})^2} $. Using the same approach, we obtain $\lambda_2 =\frac{ \sin \left( 0.5 N_\mathrm{H} \pi  ( \cos\theta_{\mathrm{B}}-\sqrt{1-(\sin{\theta_{\mathrm{B}}} \pm \frac{3}{N_\mathrm{V}})^2})\right)  }{ \sin(\frac{3 \pi}{2 N_\mathrm{V}}) \sin \left( 0.5  \pi  ( \cos\theta_{\mathrm{B}}-\sqrt{1-(\sin{\theta_{\mathrm{B}}} \pm \frac{3}{N_\mathrm{V}})^2})\right)  } $. 

\subsection{Proof of Theorem 1}	
 \label{sec:proof3}
We define the objective of \eqref{eq:max_min_P3} as a function of the parameter $M_s$ written as
\begin{align}&f(M_s) = \log_2\left( 1+\gamma_{\mathrm{B}}^\star\right) \! - \! \log_2\left( 1+ \gamma_{\mathrm{E}}^{\text{UB},1} \right) \nonumber \\
	& = \log_2\left( 1\! + \!\frac{P}{\sigma_{\mathrm{B}}^2} L_\mathrm{G}(R_1)L_\mathrm{H}(R_\mathrm{B}) \frac{(2 M_s\! - \!M)^2}{M}(2N_s\! - \!N)^2 \right) \nonumber \\ &- \log_2\left( 1+  \frac{(2M_s-M)^2(M-1)}{4M^2 M_s(M-M_s)\sin^2(\frac{3 \pi}{2 M})}\right)  \nonumber \\ 
	& = \log_2\left( 1+ \eta_\mathrm{B} (2 M_s-M)^2\right) \nonumber \\ &-\log_2\left( 1+ \eta_\mathrm{E} \frac{(2 M_s-M)^2}{4M_s(M-M_s)}\right),
\end{align}
where $\eta_\mathrm{B} = \frac{P}{\sigma_{\mathrm{B}}^2} L_\mathrm{G}(R_1)L_\mathrm{H}(R_\mathrm{B})  \frac{1}{M}(2N_s-N)^2 $ and $\eta_\mathrm{E} = \frac{(M-1)}{M^2 \sin^2(\frac{3 \pi}{2 M})-1}$  are two auxiliary parameters.  Note that $\eta_\mathrm{B}$ and $\eta_\mathrm{E}$ are independent of $M_s$. The optimal $M_s$ that maximizes $f(M_s)$ can be found by solving $\frac{\partial f(M_s)}{\partial M_s} = 0$, which is equivalent to 
\begin{align} 
	\label{eq:derivative_1}
	16 \eta_\mathrm{B} M_s^2(M\!-\!M_s)^2 \!-\!\eta_\mathrm{B}\eta_\mathrm{E} (2M_s\!-\!M)^4\!-\!M^2\eta_\mathrm{E}\!=\!0. 
\end{align}
In order to solve \eqref{eq:derivative_1}, we introduce a parameter $\tilde{M}_s = 2 M_s -M$. Thereby, we have $M_s (M-M_s) = (\frac{M}{2} +\frac{\tilde{M}_s}{2}) (\frac{M}{2} -\frac{\tilde{M}_s}{2})$. Thus, \eqref{eq:derivative_1} can be written as
\begin{align} 
	\label{eq:derivative_2}
	\eta_\mathrm{B}  (M^2-\tilde{M}_s^2)^2 -\eta_\mathrm{B}\eta_\mathrm{E} \tilde{M}_s^4-M^2\eta_\mathrm{E} =0,
\end{align}
whose solution is given by 
$
	\tilde{M}_s^* = M \sqrt{\frac{1-\frac{1}{M} \sqrt{\eta_\mathrm{B}^{-1}\eta_\mathrm{E}(\eta_\mathrm{E}-1)+M^2}}{1-\eta_\mathrm{E}}}.	
$
As a result, the optimal value of $M_s$ can be observed by $M_s^* = \frac{1}{2}(	\tilde{M}_s^*  +M)$. That completes the proof of \eqref{eq:opt_M_s}. 
By following the same approach, we obtain the optimal value of $N_s$ that maximizes the objective of \eqref{eq:max_min_P4} given by \eqref{eq:opt_N_s}.
%\textcolor{blue}{ \st{Since there is only one optimum for both $M_s$ and $N_s$, we can also confirm that the objectives of P1 and P2 are convex.}}
%\textcolor{orange}{Since the objectives of \eqref{eq:max_min_P3} and \eqref{eq:max_min_P4} both have only one zero point located in the range of $[\frac{M}{2}, M]$ and  $[\frac{N}{2}, N]$ in their first derivatives, we can confirm that they are convex. }
\bibliographystyle{IEEEbib} 
\bibliography{refs}  
\end{document}